\documentclass[11pt]{CGC2}

\usepackage[english]{babel}
\usepackage[utf8]{inputenc}
\usepackage{verbatim}
\usepackage{psfrag}
\usepackage{graphicx}
\usepackage{braket}
\usepackage{multirow}
\usepackage{color}

\usepackage{amssymb}
\usepackage{enumerate}
\usepackage{booktabs}
\usepackage{caption}
\usepackage[all]{xy}
\usepackage{mathtools}

\newcommand{\ww}{\mathrm{w}}
\newcommand{\rk}{\mathrm {rk}}

\newcommand{\BCH}{\mathrm {BCH}}
\newcommand{\HT}{\mathrm {HT}}
\newcommand{\DFT}{\mathrm {DFT}}
\newcommand{\D}{\Delta}

\newcommand{\Fq}{ \mathbb{F}_{q}}
\newcommand{\Dp}{{\D\!}^{\! +}}
\newcommand{\UU}{{\mathcal U}}

\newcommand{\vv}{\mathbf{v}}
\newcommand{\uu}{\mathbf{u}}
\newcommand{\bw}{\mathbf{w}}

\begin{document}
\Logo{BCRI--CGC--preprint, http://www.bcri.ucc.ie}

\begin{frontmatter}

\title{A generalization of bounds for cyclic codes, including the HT and BS bounds}

{\author{Matteo Piva}}
{\tt{(piva@science.unitn.it) }}\\
{{Department of Mathematics,
 University of Trento, Italy }}

 {\author{Massimiliano Sala}} {\tt{(maxsalacodes@gmail.com)}}\\
{ Department of Mathematics, University of Trento, Italy}

\runauthor{M.~Piva, M.~Sala}

\begin{abstract}
We use the algebraic structure of cyclic codes and some properties of the discrete Fourier transform to give a reformulation of several classical bounds for the distance of cyclic codes, by extending techniques of linear algebra. We propose a bound, whose computational complexity is polynomial bounded, which is a generalization of the Hartmann-Tzeng bound and the Betti-Sala bound. In the majority of computed cases, our bound is the tightest among all known polynomial-time bounds, including the Roos bound.     
\end{abstract}

\end{frontmatter}

\section*{Introduction}
\label{intro}
There are many lower bounds for the distance of cyclic codes that use some particular patterns in the set of zeros of the generator polynomial, as for example the Bose-Chaudhuri-Hockenheim ($\BCH$) bound \cite{CGC-cd-art-BCH}, the Hartmann-Tzeng ($\HT$) bound \cite{CGC-cd-art-HT}, the Betti-Sala (BS) bound \cite{CGC-cd-art-maxbetti} and the Roos bound \cite{CGC-cd-art-roos}.
We focus on this kind of bounds. To be more precise,  all bounds cited here have two important properties: their computational cost is polynomially bounded in the code length and, once any defining set of the code is given, they are independent from the code field. These bounds can be proved using an approach involving the Discrete Fourier Transform (DFT), by adapting linear algebra techniques to a set, $\UU$, not endowed with a ring structure, as shown in \cite{CGC-cd-prep-maxbetti} and \cite{CGC-cd-phdthesis-schaub}. 

We propose another bound based on the knowledge of the defining set, which is a generalization of the $\HT$ bound and the BS bound (and so it generalizes also the $\BCH$ bound) and it is independent from the Roos bound. Also our bound is polynomial time, and we call it ``bound $\mathtt{C}$''. Bound $\mathtt{C}$ follows from two partial results, bound I and bound~II, that we prove separately.  

We have run extensive computational tests. Considering all checked codes, bound $\mathtt{C}$ turns out to be the tightest among all known polynomial-time bounds.

For other polynomial-time bounds based on the structure of the defining set, you can see \cite{CGC-cd-art-ZEHWACBEZ2012} or the Carlitz-Uchiyama bound (\cite{CGC-cd-art-carlitzuchiyama}).  
We do not consider bounds which have an exponential computational cost in the code length (\cite{CGC-cd-art-vehel2}), as for example the Van Lint- Wilson bound (\cite{CGC-cd-art-vanlintd1}) or the Massey-Schaub bound (\cite{CGC-cd-phdthesis-schaub}).    

The structure of the paper is the following:
\begin{itemize}
 \item Section~\ref{Preliminaries} presents some well-known facts in literature about cyclic codes; here we recall the results concerning the use of the DFT to determine the distance of a code, which include Blahut's theorem, we provide the definition of the set $\UU$ and we explain how to use linear algebra on $\UU$. Finally we reformulate some classical bounds using the notation from $\UU$.  
 \item Section~\ref{ourbound} contains the statements and the proofs of bound I and bound~II, which imply bound $\mathtt{C}$. Our proofs rely heavily on linear algebra over $\UU$. We also show that bound $\mathtt{C}$ generalizes the $\HT$ bound and the BS bound.
 \item In Section~\ref{results} we discuss the computational complexity of bound $\mathtt{C}$ and of other classical bounds. Our bound has a complexity of $O(n^5)$. We report a tightness table that shows the behaviour of bound $\mathtt{C}$ in $1434428$ cases.
\end{itemize}


\section{Preliminaries}
\label{Preliminaries}
This part presents our notation and some preliminary results following mainly \cite{CGC-cd-book-macwilliamsI}, \cite{CGC-cd-prep-maxbetti} and \cite{CGC-cd-art-maxbetti}. 


\subsection{Backgrounds}
\label{back}
Let $n\geq 1$ and $N\geq 1$ be two natural numbers. We indicate with $(N)_n$ the remainder of the division of $N$ by $n$ and with $(N,n)$ their greatest common divisor. Let $\Fq$ be the finite field with $q$ elements, where $q$ is a power of a prime number $p$. The (Hamming) distance between two vectors $v=(v_{0},\dots, v_{n-1})$, $v'=(v'_{0},\dots, v'_{n-1})$ in $(\Fq)^n$ is the number of components for which they differ $d(v,v')=|\Set{i \mid v_{i}\neq v'_{i}, \ 0\leq i \leq n-1}|$; the (Hamming) \texttt{weight} of a vector $v=(v_0,\dots, v_{n-1})$ in $(\Fq)^n$ is the number of its non-zero coordinates:\\ $\ww(v)=|\Set{i \mid v_i\neq 0, \ 0\leq i \leq n-1}|$.
A linear code $C$ is a vector subspace of the $n$-dimensional vector space $(\Fq)^n$ and $n$ is called the \texttt{length} of $C$. \\
The \texttt{distance} of $C$ is $\min\Set{d(c_1,c_2) \mid c_1\neq c_2 \in C}=\min\Set{\ww(c)\mid 0\neq c \in C}$. If a linear code $C$ has dimension $k$ and distance $d$ we call it an $[n,k,d]$ code.

From now on, we suppose as usual $(n,q)=1$ (for the other case see~\cite{CGC-cd-art-vanlintroot}).
Let $C$ be an $[n,k,d]$ linear code over $\Fq$, $C$ is called \texttt{cyclic code} if it is an ideal of the ring $R_{n}=\Fq[x]/(x^n-1)$; a  \texttt{word} $c=(c_0,\dots,c_{n-1})\in C$ can be identified with the polynomial $c_{0}+c_{1} x+ \dots +c_{n-1}x^{n-1}$. We can characterize more precisely a cyclic code over $\Fq$: a linear code $C[n,k,d]$ is cyclic if there is a  monic polynomial $g_C\in \Fq[x]$ such that $C=\Set{g_{C} f \mid f \in \Fq[x], \ \deg{f}\leq k-1}$ and $g_{C}\mid (x^n-1)$; $g_{C}$ is called \texttt{the generator polynomial} of $C$ and it holds that $\deg(g_{C})=n-k$. Vice versa, any monic polynomial $g_{C}$ dividing $(x^n-1)$ generates a cyclic code of dimension $k=n-\deg(g_{C})$.

Let $\FF$ be the splitting field of $(x^n-1)$ over $\Fq$, i.e. $\FF=\FF_{q^m}$ where $m$ is the least positive integer such that $n\mid (q^m-1)$, and let $\alpha$ be a primitive $n$-th root of unity in $\FF$. The \texttt{(complete) defining set} of $C$ with respect to $\alpha$ is:
\begin{center}
$
S_{C}=S_{C,\alpha}=\Set{i \mid g_{C}(\alpha^i)=0, \ 0\leq i\leq n-1}. 
$
\end{center}
Let $C_j$ be the cyclotomic coset of $j$ modulo $n$, i.e. $$C_j=\Set{(j)_n, (qj)_n, (q^2j)_n, \dots, (q^{m-1}j)_n}.$$ 
If $g(\alpha^j)=0$, then $g(\alpha^{qj})=0$, thus $S_{C}$ is an union of cyclotomic cosets: $S_{C}=\cup_{j\in J} C_j$ and therefore a cyclic code is completely identified from any set $S$ containing at least one value per each cyclotomic coset, which some authors call a ``defining set''. In fact, $c\in C$ if and only if $c(\alpha^i)=0$ for any $i \in S$.


\subsection{DFT and Blahut's Theorem}
\label{DFT}
Let $\mathbb{K}$ be any field and $\alpha$ be any primitive $n$-th root of unity over $\KK$. In this general context we can define a tool we need to prove our bound: the Discrete Fourier Transform (DFT), closely related to the Mattson-Solomon polynomial.  

\begin{definition}\label{dft}
 Let $\overline{a}=(a_{0}, \dots, a_{n-1})$ be any vector over $\KK$. The \emph{Discrete Fourier Transform (DFT)} of $\overline{a}$ is the vector:
 \begin{align*}
  \DFT(\overline{a})= (A_0,\dots, A_{n-1}), && A_{i}=\sum_{j=0}^{n-1}a_{j}\alpha^{ij}
 \end{align*}
\end{definition}

\begin{remark}
 Definition~\ref{dft} is from \cite{CGC-cd-art-schaub} and it is slightly different from \cite{CGC-cd-book-macwilliamsI}, which defines $\DFT(\overline{a})=(A_{1},\dots, A_{n-1},A_{n})=(A_{1},\dots, A_{n-1},A_{0})$. We find our formulation more convenient.   
\end{remark}
\ \\
\noindent
Let $\mathbb{E}$ be the splitting field of $(x^n-1)\in \KK[x]$, then we have $\DFT(\overline{a})\in \mathbb{E}^n$. We have an isomorphism of vector spaces between $\mathbb{E}^n$ and $\mathbb{E}[x]/(x^n-1)$, which allows to identify $\DFT(\overline{a})\longleftrightarrow a_{0}+a_{1}x+\dots+a_{n-1}x^{n-1}$.
\ \\
\begin{remark}\label{remark:DFT}
 Let $C$ be a cyclic code over $\Fq$ of length $n$. We can represent a word $c\in C$ as a polynomial in $\Fq[x]\subset\FF[x]$, and then $A_i=0\iff c(\alpha^i)=0$, since $A_i=c(\alpha^i)$.  
\end{remark}
From now on $q$, $n$ and $\alpha$ are understood.
\begin{definition} Let $\overline{a}=(a_{0}, \dots, a_{n-1})$ be a vector over $\KK$. We denote by $M(\overline{a})$ the circulant matrix:
$
 \scriptstyle M(\overline{a})= 
\left(\begin{smallmatrix}
  a_{0} 		& a_{1} 	& \dots 	   	& a_{n-2} 	& a_{n-1} 	\\
  a_{n-1} 	& a_{0}	& a_{1} 		& \dots		& a_{n-2}	\\
  \vdots 	&\vdots	& \vdots		& \vdots 	& \vdots 	\\
  a_{1}		& \dots & a_{n-2}	& a_{n-1}	& a_{0}		\\
 \end{smallmatrix}\right)
 .
$
We call $M(\overline{a})$ the \texttt{matrix associated} to $\overline{a}$ and we indicate its rank as $\rk(M(\overline{a}))$.
\end{definition}
\pagebreak

We collect some elementary results of linear algebra applied to $M(\overline{a})$.
\begin{lemma}\label{lemma:LinAlg}
 Let $\overline{a}=(a_{0}, a_{1}, \dots, a_{n-1})$ be a vector in $\KK^n$.
 \begin{itemize}
  \item If $\mathrm{sh}({\overline{a}})$ is a shift of $\overline{a}$, i.e. $\mathrm{sh}(\overline{a})=(a_{n-1},a_{0}, a_{1}, \dots, a_{n-2})$, then  $$\rk(M(\overline{a}))=\rk(M(\mathrm{sh}(\overline{a}))).$$
  \item Let $\sigma$ be a permutation acting over the set $\Set{1,\dots,n}$. If $M'(\overline{a})$ is the matrix obtained by permuting the rows of $M(\overline{a})$ via $\sigma$, then   
  $$\rk(M(\overline{a}))=\rk(M'(\overline{a})).$$
  \item If $\widehat{a}$ is the reflection of $\overline{a}$, i.e. $\hat{a}=(a_{n-1},a_{n-2}, \dots, a_{1},a_{0})$, then  $$\rk(M(\overline{a}))=\rk(M(\hat{a})).$$
 \end{itemize}
\end{lemma}

We are now ready to present the main classical result in this section (see \cite{CGC-cd-prep-maxbetti}, \cite{CGC-cd-art-schaub}, \cite{CGC-cd-phdthesis-schaub}, \cite{CGC-cd-art-blahut})
\begin{theorem}[Blahut's Theorem]\label{theo:blahut}
 Let $C$ be any cyclic code, then the weight of any word $c\in C$ is equal to the rank of the matrix associated to $\DFT(c)$, i.e. $\ww(c)=\rk(M(\DFT(c))$. In particular, the distance of $C$ is:
 $$
  d=\min\Set{\rk(M(\DFT(c))) \mid c \in C, c \neq 0}.
 $$
\end{theorem}

Thus, to give estimate on the distance of a cyclic code, we can bound the rank of the matrices associated to its non-zero words; this reduces to computing the rank of the matrix associated to a vector for which \textbf{some} components are known to be zero. We formalize this information introducing the set $\mathcal{U}$ in the next section.  


\subsection{Linear algebra in the set $\UU$} 
\label{setU}
\begin{definition}
 Let $\mathcal{U}$ be a set with three symbols $\Set{0,\Dp, \D}$. We define two operations, sum and product, on $\mathcal{U}$ as follows:
\[\scriptsize{
  \begin{array}{lccccccccccccr}
 \mbox{SUM}					&&&&&&&&&&&&&	\mbox{PRODUCT}					\\
  +\colon\UU\times\UU\longrightarrow\UU	 	&&&&&&&&&&&&& 	\cdot\colon\UU\times\UU\longrightarrow\UU		\\
						 
  \begin{tabular}{l|l|l|l|}				
   $+$ 		& $0$ 	& $\D$ 	& $\Dp$	\\
   \hline
   $0$		&$0$ 	& $\D$ 	& $\Dp$	\\
   $\D$		&$\D$ 	& $\D$ 	& $\D$	\\
   $\Dp$		&$\Dp$ 	& $\D$ 	& $\D$	\\
  \end{tabular}					&&&&&&&&&&&&&

  \begin{tabular}{l|l|l|l|}
   $\cdot$ 	& $0$ 	& $\D$ 	& $\Dp$	\\
   \hline
   $0$		&$0$ 	& $0$ 	& $0$	\\
   $\D$		&$0$ 	& $\D$ 	& $\D$	\\
   $\Dp$		&$0$ 	& $\D$ 	& $\Dp$	\\
  \end{tabular}	
  \\
 \end{array}}
\]
 \end{definition}
 
 Clearly $\UU$ is not a field, but we introduce it to represent a field where we have \textbf{partial} information on the elements.
 \pagebreak

 \noindent More precisely:
 \begin{itemize}
  \item $\Dp$ represents an element of $\KK$ for which we know it is different from zero,
  \item $0$ represents an element of $\KK$ for which we know it is zero,
  \item $\D$ represents an element of $\KK$ for which we do not know if it is zero or we do not care.
 \end{itemize}
 
 \begin{example}
  Sum and product are defined over $\UU$ following the interpretation of the symbols $0$, $\D$, $\Dp$. In fact, $\Dp\cdot\Dp=\Dp$ is equivalent to saying that the product of two non-zero elements is different from zero, while $\Dp+\Dp=\D$ is equivalent to saying that the sum of two non-zero elements could be zero or non-zero.     
 \end{example}

Although $\UU$ is not a field and $\UU^n$ is not a vector space, it is useful to adopt for them some terms from the theory of vector spaces.

\begin{definition}
Let $\uu=(u_{0}, \dots , u_{n-1})$ be an element of $\UU^n$. We call $\uu$ a \texttt{vector} of $\UU^n$ and we also write $\uu[i]=u_{i-1}$ for any $i\in\Set{1,\dots n}$.
\end{definition}
\begin{remark}\label{rem:vectU}
 Let $k\in\ZZ$ be any integer and $\uu\in \UU^n$. For convenience, sometimes we write $\uu[k]$, meaning:
 \[
  \uu[k]=
  \begin{cases}
    \uu[(k)_n]	&	\text{ if $(k)_ n \neq 0$} \\
    \uu[n]	&	\text{ otherwise.}
  \end{cases}	
 \]

\end{remark}

\begin{definition}
We indicate with $M(\mathbf{u})\in\UU^{n\times n}$ the circulant matrix obtained from a vector $\uu$ in $\UU^n$.
\end{definition}
We say that a set of vectors is linear independent in $\UU^n$ if they correspond to a set of linear independent vectors in every vector space $\KK^n$. To define this notion in a way useful for our proofs, we need a couple of definitions.
\begin{definition}\label{def:instance}
Let $n\geq 1$ be a natural number, $\uu=(u_0,\dots,u_{n-1})\in\UU^n$. An \texttt{instance} of $\uu$ over $\KK$ is any vector $v=(v_{0},\dots,v_{n-1}) \in \KK^n$ such that for $0\leq i \leq n-1$:
\begin{enumerate}
 \item $ v_i=0$ if $u_i=0$,
 \item $ v_i \neq 0$ if $u_i=\Dp.$
\end{enumerate}
  The set of all instances of $\uu$ is called \texttt{instantiation} of $\uu$ over $\KK$ and we write $\mathtt{In}(\uu,\KK)=\Set{v \in \KK^n \mid v \mbox{ is an instance of } \uu}$.
\end{definition}

\begin{remark}
 Note that in Definition~\ref{def:instance} we did not specify the value of $v_i$ when $u_i=\D$, so $v_i$ can be freely chosen for this value of $i$.
\end{remark}
\pagebreak

\begin{example} Let us consider $\KK=\FF_2$. 
\begin{itemize}
 \item[-] if $\uu=(0,\D,\Dp)\in \UU^3$, then $\mathtt{In}(\uu,\FF_2)=\Set{(0,0,1),(0,1,1)}$
 \item[-] if $\uu=(0,\Dp,\D)\in \UU^3$, then $\mathtt{In}(\uu,\FF_2)=\Set{(0,1,0),(0,1,1)}$.
\end{itemize}
\end{example}

\begin{definition}\label{def:linInd}
 Let $s\geq 1$. We say that $\uu^{1},\ldots,\uu^{s}\in \UU^n$ are \texttt{linear independent} if for any field $\KK$, for any $v^i \in \mathtt{In}(\uu^i,\KK)$ with $1\leq i\leq s$,
 we have that $\Set{v^i}_{1\leq i \leq s}$ are linear independent (over $\KK$).
\end{definition}

In other words, for any instance set $\Set{v^1,\dots,v^s}$, for any $\Set{\lambda_i}_{1\leq i\leq s}\subset\KK$:
$$
\sum_{i=1}^{s}\lambda_i v^i=0\iff \lambda_1=\dots=\lambda_s=0.
$$
 
To check if a set of vectors in $\UU^n$ is linearly independent, we use the so-called ``singleton procedure'' (see \cite{CGC-cd-art-maxbetti}, \cite{CGC-cd-prep-maxbetti}, \cite{CGC-cd-prep-maxponchio}).\\
For any matrix $M$, $M[i,j]$ is the  $(i,j)$ entry, $M[i]$ is the $i-$th row and $M(j)$ is the $j-$th column.
\begin{definition}
 Let $M$ be a matrix over $\UU$. We say that a column $M(j)$ is a \texttt{singleton} if it contains only one non-zero component $M[i,j]$, i.e. $M[i,j]=\Dp$ and $M[t,j]=0$ for $t\neq i$. When this happens we say that $M[i]$ is \texttt{the row corresponding to the singleton}. 
\end{definition}
Any ordered multiset of $t$ rows of length $n$ with $t\leq n$ forms a matrix $M_t \in \UU^{t \times n}$. If a column $M(j)$ is a singleton, then the row corresponding to the singleton is clearly linear independent (in $\UU$) from the others. Then we can delete the $j-th$ column and the corresponding row (we call this operation \texttt{s-deletion}), obtaining a new matrix, $M_{t-1}$, and we search for a new singleton in $M_{t-1}$. If this procedure can continue until we find a matrix $M_1$ with at least one $\Dp$, we say that \texttt{the singleton procedure is successful} for the set of $t$ rows considered.  
\begin{definition}\label{def:prk}
 Let $M$ be a matrix over $\UU$, we denote by $\rk(M)$ the \texttt{rank} of $M$ and by $\mathrm{prk}(M)$ the \texttt{pseudo-rank} of $M$, i.e. 
 \begin{itemize}
  \item the rank $\rk(M)$ is the largest $t$ such that there exists a set of $t$ rows in $M$ which are linearly independent
  \item the pseudo-rank $\mathrm{prk}(M)$ is the largest $t$ such that there exists a set of $t$ rows in $M$ for which the singleton procedure is successful.
 \end{itemize}
\end{definition}

Clearly, $\rk(M)\geq\mathrm{prk}(M)$. We collect in the following lemma some elementary results for rank and  pseudo-rank of matrices over $\UU$.
\pagebreak

\begin{lemma}\label{lemma:LinAlgU}
 Let $\uu$ be a vector in $\UU^n$.
 \begin{itemize}
  \item If $\vv$ is a obtained from a shift of $\uu$ by any number of positions, then  
  \begin{align*}
\rk(M({\bf v}))=\rk(M({\bf u}))	&&	\mathrm{prk}(M({\bf v}))=\mathrm{prk}(M({\bf u})).   
  \end{align*}

  \item Let $\sigma$ be a permutation in the symmetric group $S_n$. Let $M \in {\mathcal
    U}^{n \times n}$ and $M'$ be the matrix obtained by applying $\sigma$ to the rows of $M$. Then 
      \begin{align*}
       \rk(M(\uu))=\rk(M'(\uu))	&&	\mathrm{prk}(M(\uu))=\mathrm{prk}(M'(\uu)).
      \end{align*}

  \item If $\hat{\bf u}$ is the reflection of $\uu$, i.e. the vector in ${\mathcal U}$ s.t. ${\bf u}[i]=\hat{\bf u}[n-i+1]$ for any $1\leq i\leq n$, then
     \begin{align*}
      \rk(M({\bf u}))=\rk(M(\hat{\bf u}))	&&	\mathrm{prk}(M({\bf u}))=\mathrm{prk}(M(\hat{\bf u})).
     \end{align*}

 \end{itemize}
\begin{proof}
 The equalities regarding the rank follow from Lemma~\ref{lemma:LinAlg}. The equalities regarding the pseudo-rank admit easy proof, that we omit.
\end{proof}

\end{lemma}

Given any cyclic code $C$ of length $n$ and defining set $S_C$, there is a natural way to see $S_C$ as a vector in $\UU^n$.

\begin{definition}
 Let $C$ be a cyclic code of length $n$ with defining set $S_C$. We denote with $R(n, S_C)$ the vector $(u_{0},\dots,u_{n-1})\in \UU^n$ such that for $0\leq i \leq n-1$:
 \[
  u_i=
  \begin{cases}
   0,	&	 \text{if $i \in S_C$}	\\
   \D, 	&	 \text{otherwise.}
  \end{cases}
 \]
\end{definition}

\begin{definition}\label{def:setA}
 Given $\vv\in \UU^n$, $\mathcal{A}(\vv)$ is the set of all $\mathbf{u}\in\UU^n\setminus\mathbf{0}$ s.t.
\begin{align*}
 & 	\mathbf{u}[i]=0,  \mbox{ if } \vv[i]=0, 					\\
 &	\mathbf{u}[i]=\Dp,  \mbox{ if } \vv[i]=\Dp, 					\\
 &	\mathbf{u}[i]=\Dp  \mbox{ or } \mathbf{u}[i]=0, \mbox{ if } \vv[i]=\D. 	\\
\end{align*}
\end{definition}

Our interest for the rank of a matrix on $\UU$ is due to the following result.
\begin{theorem}\label{th:Schaub}
 Let $C$ be a cyclic code with defining set $S_C$ and length $n$. If $d$ is the distance of the code, then
 \begin{align*}
  d &	\geq\min\Set{\rk(M(\mathbf{u})) \mid \mathbf{u} \in \mathcal{A}(R(n,S_C))}		\\
    &	\geq\min\Set{\mathrm{prk}(M(\mathbf{u})) \mid \mathbf{u} \in \mathcal{A}(R(n,S_C))}	\\
 \end{align*}

\begin{proof}
 See \cite{CGC-cd-phdthesis-schaub} or \cite{CGC-cd-prep-maxponchio}
\end{proof}
\end{theorem}


\subsection{Bounds and $\UU$}
\label{BoundU}
Many classical bounds for the distance of cyclic codes are based on the knowledge of the defining set. In this section we provide the statements of some classical bounds using the notation induced by the set $\UU$.\\       
Let $i\geq 1$. We define three patterns of symbols, which we call ``blocks'':
\begin{align*}
 (0)^i=(\overbrace{0,\dots,0}^{i}), 	&&  (\D)^i=(\overbrace{\D,\dots,\D}^{i}), 	&& (\Dp)^i=(\overbrace{\Dp,\dots,\Dp}^{i}),	\\
																	\\
					&&	(0)^0=(\D)^0=(\Dp)^0=\emptyset. 		&&
\end{align*}

Using these three first blocks we can define multiple blocks using concatenation, for example $(0)^3(\D)^2=(0,0,0,\D,\D)$ or $(0)^2(\Dp)^4=(0,0,\Dp,\Dp,\Dp,\Dp)$. We also define blocks of blocks, with an obvious meaning, as for example:
\[
 ((0)^2(\Dp)^3)^2(\D)^2=(0,0,\Dp,\Dp,\Dp,0,0,\Dp,\Dp,\Dp,\D,\D).
\]
Let us consider two vectors of different length, for example: 
\begin{align*}
\uu=(\Dp,0,\D)\in\UU^3, && \vv=(\Dp,0,\D,\Dp,\Dp,0)\in\UU^6 .
\end{align*}
Let $\KK$ be any field, then the vector $\uu$ represents a vector in $\KK^3$ with the first coordinate different from zero, the second coordinate equal to zero and the third component that is any element of $\KK$. In the same way, $\vv$ represents a vector of $\KK^6$ such that the first, the fourth and the fifth component are different from zero, the second component is zero and the third component is any element of $\KK$. \\
We note that the constraints for the components of $\uu$ coincide with the constraints for the first three components of $\vv$ and in this case we write  $\uu\preccurlyeq\vv$.
The previous example shows a particular case of a special kind of relation among vectors over $\UU$, that we are going to define in the following definition. 
%
\begin{definition}\label{def:contained}
Let $n,m \in \NN$ such that $n\geq m$. Let $\pi$ be the projection of $\UU^n$ on $\UU^m$ as follows:
\begin{equation*}
 \pi\colon\UU^n\to\UU^m ,\qquad \pi((v_1,\dots,v_n))=(v_1,\dots,v_m).
\end{equation*}
Let $\uu\in\UU^m$ and $\vv\in\UU^n$, we write $\uu \preccurlyeq \vv$ if there is $0\leq i \leq n-1$ such that 
$$
\mathcal{A}(\pi(\mathrm{sh}^i(\vv))\subseteq\mathcal{A}(\uu).                                                                                                                                                                                                                                                            $$
When $\uu\preccurlyeq\vv$ we say that $\uu$ is \texttt{included} in $\vv$.
\end{definition}
Our Definition~\ref{def:contained} of inclusion of vectors has some particular properties that we are going to show.
\begin{proposition}\label{prop:properties}
Let $\uu\in\UU^m$, $\vv\in\UU^n$, $\bw\in\UU^t$ with $m,n,t\geq 1$. We indicate with $\uu\vv$ the vector in $\UU^{m+n}$ obtained by concatenating $\uu$ and $\vv$, i.e. $\uu\vv=(u_1,\dots,u_m,v_1,\dots,v_n)$. The following statements hold:
 \begin{enumerate}[a)]
  \item \label{prop:a}$(\D) \preccurlyeq (\Dp)$, $(\Dp) \not\preccurlyeq (\D)$, $(\D) \preccurlyeq (0)$, $(\D) \not\preccurlyeq (0)$.
  \item \label{prop:c}$ \vv \preccurlyeq \vv$.
  \item  \label{prop:b}$\uu \preccurlyeq \vv \iff \uu \preccurlyeq \mathrm{sh}(\vv)$.
  \item \label{prop:d}$\vv \preccurlyeq \uu\vv$, $\vv \preccurlyeq \vv\uu$. 
  \item \label{prop:e}$\vv \preccurlyeq \uu\vv\bw$.   
  \item \label{prop:f}$(\D)^m \preccurlyeq \vv$ for any $\vv\in \UU^n$ s.t. $m\leq n$.
 \end{enumerate}
\begin{proof} \
\begin{enumerate}[a)]
  \item Since $(\D),(\Dp),(0)\in\UU^1$ the shift is trivial and then we can ignore it. We have: $\mathcal{A}((\D))=\Set{(\Dp)}$, $\mathcal{A}((\Dp))=\Set{(\Dp)}$, $\mathcal{A}((0))=\emptyset$,
  \begin{align*}
   \mathcal{A}((\Dp))\subseteq\mathcal{A}((\D)),	&&	\mathcal{A}((\D))\not\subseteq\mathcal{A}((\Dp)),	\\
   \mathcal{A}((0))\subseteq\mathcal{A}((\D)),		&&	\mathcal{A}((\D))\not\subseteq\mathcal{A}((0)).
  \end{align*}
  \item Since $n=m$ the projection becomes trivially the identity and it is sufficient to take $i=0$ in order to have $\mathcal{A}(\pi(\mathrm{sh}^0(\vv)))=\mathcal{A}((\vv))\subseteq\mathcal{A}(\vv)$.
  \item ``$\implies$''. Let $\overline{\vv}=\mathrm{sh}(\vv)$ and let $0\leq i\leq n-1$ be s.t. $\mathcal{A}(\pi(\mathrm{sh}^i(\vv)))\subseteq\mathcal{A}(\uu)$. 
  Denoting $\overline{i}=(i-1)_{n}$ we have $\mathrm{sh}^{\overline{i}}(\overline{\vv})=\mathrm{sh}^{i}(\vv)$ and so $\mathcal{A}(\pi(\mathrm{sh}^{\overline{i}}(\overline{\vv})))\subseteq\mathcal{A}(\uu)$ which implies $\uu\preccurlyeq\overline{\vv}$. The proof of ``$\impliedby$'' is analogous.
  \item Since $\mathrm{sh}^{n}(\uu\vv)=\vv\uu$, we have $\pi(\vv\uu)=\vv$, $\mathcal{A}(\pi(\mathrm{sh}^{n}(\uu\vv)))=\mathcal{A}(\pi(\vv\uu))=\mathcal{A}(\vv)\subseteq\mathcal{A}(\vv)$.
  In the same way $\pi(\vv\uu)=\vv$ and $\mathcal{A}(\pi(\vv\uu))=\mathcal{A}(\vv)\subseteq\mathcal{A}(\vv)$.
  \item From (\ref{prop:d}) we have that $\vv\preccurlyeq\vv\bw\uu$ for all $\bw\uu\in\UU^{t+m}$ and since $\uu\vv\bw=\mathrm{sh}^{m}(\vv\bw\uu)$ we use (\ref{prop:b}) to conclude that $\vv\preccurlyeq\uu\vv\bw$.
  \item We have $\mathcal{A}((\D)^m)=\Set{0,\Dp}^m\setminus \mathbf{0}$ and by Definition~\ref{def:setA} $\mathcal{A}(\pi(\vv))\subseteq\Set{0,\Dp}^m\setminus\mathbf{0}$ for any $\vv\in\UU^n$, $m\leq n$.
 \end{enumerate}
 \end{proof}
\end{proposition}
\begin{example}\
\begin{itemize}
 \item $(\D,\Dp\D)\preccurlyeq(0,0,\D,\Dp,\D,\Dp)$ by Proposition~\ref{prop:properties} - (\ref{prop:e}), since $(\D,\Dp\D)\preccurlyeq(0,0)(\D,\Dp,\D)(\Dp)$;
 \item $(0)^2(\D)\preccurlyeq (0,\D,\Dp,\D,0)$ by Proposition~\ref{prop:properties} - (\ref{prop:b})-(\ref{prop:d}), since $(0)^2(\D)\preccurlyeq (0,0)(\D,\Dp,\D),$ and we can obtain $(0,\D,\Dp,\D,0)$ if we shift by $n-1$ positions;
 \item $(0,\Dp,\Dp)\not\preccurlyeq(\Dp,\Dp,0,\D,\D),$ because we have $\mathcal{A}\left((0,\Dp,\Dp)\right)=\Set{(0,\Dp,\Dp)}$, and for $0\leq i\leq 4$:
 \[ 
  i=0	\ \ \ \mathcal{A}(\pi(\Dp,\Dp,0,\D,\D))=\Set{(\Dp,\Dp,0)}
  \]
  \[
  i=1	\ \ \	\mathcal{A}(\pi(\D,\Dp,\Dp,0,\D))=\Set{(\Dp,\Dp,\Dp),(0,\Dp,\Dp)}			\]
  \[
  i=2	\ \ \ \mathcal{A}(\pi(\D,\D,\Dp,\Dp,0))=\Set{(\Dp,\Dp,\Dp),(0,0,\Dp),(0,\Dp,\Dp),(\Dp,0,\Dp)}
  \]
  \[
  i=3	\ \ \ \mathcal{A}(\pi(0,\D,\D,\Dp,\Dp))=\Set{(0,\Dp,\Dp),(0,0,\Dp),(0,\Dp,0)}		
  \]
  \[
  i=4	\ \ \	\mathcal{A}(\pi(\Dp,0,\D,\D,\Dp))=\Set{(\Dp,0,0),(\Dp,0,\Dp)}	
  \]
  and then for any $0\leq i \leq 4$, $\mathcal{A}\left(\pi(\mathrm{sh}^i((\Dp,\Dp,0,\D,\D)))\right)\not\subseteq\Set{(0,\Dp,\Dp)}$;
  
 \item $(0,\Dp,\Dp)\not\preccurlyeq (\D,\Dp,0,0,\Dp)$, it is sufficient to note that it is impossible to find in $(\D,\Dp,0,0,\Dp)$ three consecutive components such that first is zero and the others are different from zero. 
\end{itemize}
\end{example}
\begin{remark}
 Proposition~\ref{prop:properties} - (\ref{prop:c}) proves that $\preccurlyeq$ is a reflexive relation. Unfortunately, it is not transitive in fact $(\Dp,0,\Dp)\preccurlyeq(\Dp,\Dp,0)$ and $(\Dp,\Dp,0)\preccurlyeq(\Dp,\Dp,0,0,\Dp)$ but $(\Dp,0,\Dp)\not\preccurlyeq(\Dp,\Dp,0,0,\Dp)$.
\end{remark}
We provide the classical definitions of three bounds which generalize the BCH bound (\cite{CGC-cd-art-BCH},\cite{CGC-cd-art-BCH2}), followed by their interpretation in $\UU$.\\
We will take for granted that $\alpha $ is an $n$-th primitive root of unity over $\FF_q$, that $C$ is an $[n,k,d]$ cyclic code over $\FF_q$ with generator polynomial $g$ and $S_C$ is the defining set.
\begin{theorem}[Hartmann-Tzeng bound, \cite{CGC-cd-art-HT}]\label{th:HT}
 Suppose that there exist $i_0,m, s,r \in \NN $ s.t. $m\geq 1$, $s\geq 1$, $(m+r,n)=1  $ for which
$$
  g(\alpha^{i_0+i +j (m+r)})=0, \quad 0\leq i\leq m-1 , \ 0 \leq j \leq s-1\,.
$$
Then
$$
d\geq m+s .
$$
\end{theorem}

In \cite{CGC-cd-art-roos2}, Roos improves the original Hartmann-Tzeng bound substituting the condition $(m+r,n)=1$ in Theorem~\ref{th:HT} with the less restrictive $(m+r,n)\leq m$. The following theorem is the Roos version of the Hartmann-Tzeng bound, written using $\UU$.
\begin{theorem}[Hartmann-Tzeng bound, \cite{CGC-cd-prep-maxbetti}]\label{th:HTU}
Suppose that there are  $m,s, r, \rho\in \NN$, $m\geq 1$, $s\geq 1$, $\rho\geq 1$, such that $(m+r,n)\leq m$ and
for which 
$$
((0^{m})(\D^{r}))^s\preccurlyeq  R(n,S_C)^\rho.
$$
Then
$$
d\geq m+s .
$$
\end{theorem}
\begin{theorem}[Betti-Sala bound, \cite{CGC-cd-art-maxbetti}]
Suppose that there are $\lambda,\mu \in \NN$, $\lambda,\mu\geq1$ and $i_0 \in \{0, \dots , n-1\}$ 
such that
\begin{align*}
&	\mbox{either}			&					&						&	\\
&					&	(i_0+j)_n \in S_C \ \ \ \ 	& j =0, \dots ,\lambda\mu-1				&	\\
&					&					&						&	\\
&					&	(i_0+j)_n \in S_C,\ \ \ \	& j=  (\lambda+h)\mu +1, \dots ,(\lambda+h)\mu+ \mu-1,	&	\\
&					&					& 0 \leq h \leq \lambda				&	\\
&	\mbox{or}			&					&						&	\\
&					&	(i_0+j)_n \in S_C \ \ \ \ 	& j=  h\mu , \dots ,h\mu+ \mu-2		&	\\
&					&					& 0 \leq h \leq \lambda				&	\\
&					&					&						&	\\
&					&	(i_0+j)_n \in S_C,\ \ \ \	& j =(\lambda+1)\mu, \dots ,(2\lambda+1)\mu-1. 		&
\end{align*}
%
%
%
%
Then:
$$d \geq \lambda\mu + \mu \,.$$
\end{theorem}
\begin{theorem}[Betti-Sala bound, \cite{CGC-cd-prep-maxbetti}]\label{th:BSU}
Suppose that there are $\lambda,\mu \in \NN$, $\lambda,\mu\geq1$ such that either
\begin{align*}
&	\mbox{either}			&												\\			
&					&	((0)^\mu)^\lambda ((\D)^1(0)^{\mu-1})^{\lambda+1} \preccurlyeq R(n,S_C)^\rho					\\
&					&												\\
&	\mbox{or}			&												\\
&					&	((0)^{\mu-1}(\D)^1)^{\lambda+1} ((0)^{\mu})^{\lambda} \preccurlyeq R(n,S_C)^\rho.			
\end{align*}
Then:
$$d \geq \lambda\mu + \mu.$$
\end{theorem}

\begin{theorem}[Roos bound, \cite{CGC-cd-art-roos}]
Let $m,r\in \NN$ s.t. $1\leq m\leq n-1$, $1\leq m+r\leq n-1$ and\footnote{$\alpha^{(m+r)}$ is another primitive n-th root of unity} $(m+r,n)=1$.
Let $\bar{S}$ be a set of $\bar s$ consecutive natural numbers:
$\bar{S}:=\{k,k+1,\ldots,k+\bar{s}-1\}$.
Let $S'\subset \bar{S}$, $|S'|=s$, s.t.\footnote{i.e., $S'$ is obtained
from $S$ by removing strictly less than $m$ elements, that we will informally call ``holes'' in Example~\ref{example:roos}}
$$
          \bar{s}-s\leq m-1 \,.
$$
Suppose that, for an $0\leq i_0\leq n-1$, we have
$$
      g(\alpha^{i_0+i+\sigma (m+r)}),\quad \mbox{ for } 
      0\leq i\leq m-1\; \mbox{and}\; \sigma\in S' \,.
$$
Then 
$$
d\geq m + s
$$  
\end{theorem}

\begin{theorem}[Roos bound, \cite{CGC-cd-prep-maxbetti}]\label{th:RoosU}
Suppose that there are $m$, $r$, $s \in \NN$, $m\geq 1$, $(m+r,n)=1 $, and there exist $s$ integers $0=k_1<k_2<..<k_s<m+s-1$, so 
that: 
\[\scriptstyle
  (\D)^{(m+r)k_1}  (0)^m (\D)^{r}(\D)^{(m+r)(k_2-k_1-1)} (0)^m(\D)^{r} \cdots 
  (\D)^{(m+r)(k_s-k_{s-1}-1)}  (0)^m (\D)^{r} \,
           \preccurlyeq  \, R(n,S)^\rho \,.
\]
Then:
$$d \geq m+s.$$
\end{theorem}
Usually, the Roos bound is presented as a generalization of the Hartmann-Tzeng bound. This is certainly true if we refer to the classical version of the Hartmann-Tzeng bound (\cite{CGC-cd-art-HT}), but it may be false if we refer to the version of the Hartmann-Tzeng bound improved by Roos (\cite{CGC-cd-art-roos2}). In fact, it may be possible to find codes for which the generalized Hartmann-Tzeng bound is sharper and tighter than Roos's.
%
%
We note that in the statements of classical bounds with our notation (Theorem~\ref{th:HTU}, \ref{th:BSU}, \ref{th:RoosU}) we use the letter $\rho$ for a special role. In these statements we are looking for special patterns in $R(n,S)$, which is a sort of $\UU$-translation of the defining set, with the sought-after pattern playing the role of the root positions. However, in classical statements the root positions are intrinsically given modulo the length of the code and so it may happen that the corresponding pattern will need two or more consecutive $R(n,S)$ sets to be matched. The maximum value of
the needed $\rho$ can be easily computed and we provide it without a proof.

\begin{fact}
The maximum value for $\rho$ is 
\begin{itemize}
 \item $\left\lfloor\frac{s(m+r)}{n}\right\rfloor+1$ for Theorem~\ref{th:HTU}
 \item $\left\lfloor\frac{\mu m+1+\mu(m+1)}{n}\right\rfloor+1$ for Theorem~\ref{th:BSU}
 \item $\left\lfloor\frac{(m+s-1)(m+r)}{n}\right\rfloor+1$ for Theorem~\ref{th:RoosU}.
\end{itemize}
\end{fact}
We now discuss an example present in \cite{CGC-cd-art-roos}, where we show how Roos's bound itself can be applied with our notation. Note in particular that here we do need $\rho>1$ and also
that we can actually increase the estimated distance, from the value provided in \cite{CGC-cd-art-roos} to the actual distance, \underline{still using} only the Roos bound. 
\begin{example}\label{example:roos}
 Let $n=21$, $q=2$ and $C$ be the cyclic code with generator polynomial $g=x^{14} + x^{13} + x^9 + x^8 + x^7 + x^5 + x^4 + x^3 + 1$. It has defining set $S_C=C_1 \cup C_3 \cup C_7 \cup C_9$, where
 \begin{align*}
  &	C_1=\Set{1,2,4,8,11,16},	&	&	C_3=\Set{3,6,12},	&	\\
  &	C_7=\Set{7,14},			&	&	C_9=\Set{9,15,18}.	&
 \end{align*}
 Then $R(n, S_C)=( \D, 0, 0, 0, 0, \D, 0, 0, 0, 0, \D, 0, 0, \D, 0, 0, 0, \D, 0, \D, \D )$. \\
 We indicate with $s'$ when we find a block ($s$ is the number of all blocks) and with $h'$ when we find a hole, recalling that the numbers of holes, $h$, must be strictly less than the numbers of zeros, $m$, in a block. 
Taking $m=3$ (so $h<3$), $r=1$ and starting from $i_0=3$ (which is equivalent to what done in \cite{CGC-cd-art-roos}, Example~1) we have:
\begin{equation*}
\begin{xy}
,(0,0)*{\D},(6,0)*{0}, (12,0)*{0}, (18,0)*{0}, (24,0)*{0}, (30,0)*{\D}, (36,0)*{0},(42,0)*{0}, (48,0)*{0}, (54,0)*{0}, (60,0)*{\D}, (66,0)*{0}, (72,0)*{0},(78,0)*{\D}, (84,0)*{0}, (90,0)*{0}, (96,0)*{0}, (102,0)*{\D}, (108,0)*{0}, (114,0)*{\D}, (120,0)*{\D};
,(21,-8)*{s'=1};
,(9.5,-2);(9.5,-4)**\dir{-};
,(9.5,-4);(32.5,-4)**\dir{-};
(32.5,-2);(32.5,-4)**\dir{-};
,(45,-8)*{s'=2};
,(33.5,-2);(33.5,-4)**\dir{-};
,(33.5,-4);(56.5,-4)**\dir{-};
,(56.5,-2);(56.5,-4)**\dir{-};
,(69,-8)*{h'=1};
,(57.5,-2);(57.5,-4)**\dir{-};
,(57.5,-4);(80.5,-4)**\dir{-};
,(80.5,-2);(80.5,-4)**\dir{-};
,(93,-8)*{s'=3};
,(81.5,-2);(81.5,-4)**\dir{-};
,(81.5,-4);(104.5,-4)**\dir{-};
,(104.5,-2);(104.5,-4)**\dir{-};
,(117,-8)*{h'=2};
,(105.5,-2);(105.5,-4)**\dir{-};
,(105.5,-4);(122.5,-4)**\dir{-};
,(-2.5,-4);(2.5,-4)**\dir{-};
,(2.5,-2);(2.5,-4)**\dir{-};
,(15,8)*{s'=s=4};
,(3.5,2);(3.5,4)**\dir{-};
,(3.5,4);(26.5,4)**\dir{-};
,(26.5,2);(26.5,4)**\dir{-}
,(39,8)*{\mathbf{h'=h=3}};
,(27.5,2);(27.5,4)**\dir{-};
,(27.5,4);(50.5,4)**\dir{-};
,(50.5,2);(50.5,4)**\dir{-}
\end{xy}
\end{equation*}
which, using $\rho=\left\lfloor\frac{(3+4-1)(3+1)}{21}\right\rfloor+1=2$ becomes:
\[
R(n,S_C)^2= \D0\underbracket[0.5pt]{000\D}_{s=1}\underbracket[0.5pt]{0000}_{s=2}\underbracket[0.5pt]{\D00\D}_{h=1}\underbracket[0.5pt]{000\D}_{s=3}\underbracket[0.5pt]{0\D\D\vline height 12pt depth 1.5pt width 1pt\D}_{h=2} \underbracket[0.5pt]{0000}_{s=4}\underbracket[0.5pt]{\D000}_{\mathbf{h=3}}0\D00\D000\D0\D\D
\]
We stopped to search since we exceed the number of admitted holes, so we find, as in \cite{CGC-cd-art-roos} that the distance of the code is at least $7$. Actually, a more detailed search reveals that the Roos bound applied with parameters $m= 2$, $r=17$ and $i_0=7$, gives distance at least $8$ which turns out to be the true distance of the code. 
\end{example}
\indent 


\section{Our bound}
\label{ourbound}
The results contained in this section appear here in full for the first time (but see \cite{CGC-alg-tesi2-piva10} for a preliminary version). As we sketched at the end of Section~\ref{DFT}, we give estimates for the rank of matrices over $\UU$, in order to obtain estimates on the distance of a cyclic codes. Our main tools are Theorem~\ref{th:Schaub} and the singleton procedure.


\subsection{Statement of bound I and bound II}
\label{bounAB}

In this section we present two propositions that compose the main result of this paper; the next section is dedicate to their proof. 
\begin{proposition}[bound I]\label{boundD1}
 Let $C$ be an $\Fq[n,k,d]$ cyclic code with defining set $S_C$ and $(q,n)=1$. Suppose that there are $\ell, \ m, \ r, \ s \in\mathbb{N} $, $1\leq m\leq \ell$ and $i_0 \in \Set{0,\dots,n-1}$ such that:
\begin{itemize}
 \item[a)] $\left(i_0+j\right)_n \in S_C$, $\forall j=0,\dots,\ell-1$,
 \item[b)]$\left(i_0+j\right)_n \in S_C,$
\begin{align} 
 &\forall j=i_0+\ell+r+h(m+r)+1,\dots, \ i_0+\ell+r+m+h(m+r)\notag\\
 &\forall 0\leq h\leq s-1\notag
\end{align}
\end{itemize} Then
\begin{itemize}
 \item if $(m+r,n)\leq m$:
 \begin{equation}\label{eq1}
 d\geq \ell +1 +s-r \left\lfloor\frac{\ell}{m+r}\right\rfloor-\max\Set{(\ell)_{m+r}-m,0};
\end{equation}
 \item otherwise
 \begin{equation}
 d\geq \ell +1 .
\end{equation}
\end{itemize}
\end{proposition}
The above statement is expressed in classical notation and seems extremely complicated. However it is a natural generalization of known bounds, as it is immediate once it is expressed in $\UU$ notation.
\begin{proposition}[bound I]\label{boundD1U}
Let $C$ be an $[n,k,d]$ cyclic code with defining set $S_C$.
Suppose that there are  $\ell,s,m,r, \rho\in \NN$, $\ell\geq m\geq 1$, $s\geq 1$, $\rho\geq 1$, $r\geq 1$ such that 
\begin{equation}\label{eq:bI}
((0^{\ell})(\D^{r}))((0^{m})(\D^{r}))^s\preccurlyeq  R(n,S_C)^\rho.
\end{equation}
Then
\begin{itemize}
 \item if $(m+r,n)\leq m$:
 \begin{equation}\label{eq1UU}
 d\geq \ell +1 +s-r \left\lfloor\frac{\ell}{m+r}\right\rfloor-\max\Set{(\ell)_{m+r}-m,0};
\end{equation}
 \item otherwise
 \begin{equation}\label{eq1UUU}
 d\geq \ell +1 .
\end{equation}
\end{itemize}
\end{proposition}
\begin{corollary}
In Proposition~\ref{boundD1U} we can substitute condition \eqref{eq:bI} with
\[
((\D^{r})(0^{m}))^s ((\D^{r})(0^{\ell}))\preccurlyeq  R(n,S_C)^\rho.
\]
\begin{proof}
 See Lemma~\ref{lemma:LinAlgU}.
\end{proof}
\end{corollary}
\begin{remark}\label{rem:D1}
We can see Proposition~\ref{boundD1} as a generalization of the $\HT$ bound. In fact with $\ell=m$ the statement of Proposition~\ref{boundD1U}-\eqref{eq1UU} reduces to Theorem~\ref{th:HTU}.
\end{remark}

We are able to prove another bound, similar to the bound I:
\begin{proposition}[bound II]\label{boundD2}
Let  $C$ be an $[n,k,d]$ cyclic code over $\FF_q$ with defining set $S_C$.
Suppose that there are $\lambda,\mu,s \in \NN$, $\lambda\geq 1$, $\mu\geq 2$, $s\geq \lambda+1$, $(n,\mu)\leq\mu-1$, $i_0 \in \{0, \dots , n-1\}$ 
such that:
\begin{enumerate}

\item[a)] $ (i_0+j)_n \in S_C$, $j =0, \dots ,\lambda\mu-1 $,

\item[b)] $ (i_0+j)_n \in S_C$, 
          $j=  (\lambda+h)\mu +1, \dots ,(\lambda+h)\mu+ \mu-1$, $ 0 \leq h \leq s-1$,
\end{enumerate}
Then:
\begin{itemize}
 \item if $(n,\mu)\leq\mu-1$:
 \begin{equation*}\label{eq2}
d \geq \lambda\mu + \mu +s-\lambda-1;
\end{equation*}
 \item otherwise if $\mu \mid n$:
 \begin{equation*}
d \geq \lambda\mu + \mu .
\end{equation*}
 
\end{itemize}

\end{proposition}
Again, the $\UU$ notation is more clear, as follows.

\begin{proposition}\label{boundD2U}
Let  $C$ be an $[n,k,d]$ cyclic code over $\FF_q$ with defining set $S_C$.
Suppose that there are $\lambda,\mu,s \in \NN$, $\lambda\geq 1$, $\mu\geq 2$, $s\geq \lambda+1$ such that:
\begin{equation}\label{eq:bII}
 (0^{\mu \lambda}\D)(0^{\mu-1}\D)^s \preccurlyeq R(n,S_C)^\rho.
\end{equation}
Then:
\begin{itemize}
 \item if $(n,\mu)\leq\mu-1$:
\begin{equation}\label{eq2UU}
 d \geq \lambda\mu + \mu +s-\lambda-1;
\end{equation}
 \item otherwise if $\mu \mid n$:
 \begin{equation}\label{eq2UUU}
 d \geq \lambda\mu + \mu .
\end{equation}
\end{itemize}
\end{proposition}
\begin{corollary}
In Proposition~\ref{boundD2U} we can substitute condition \eqref{eq:bII} with
\[
(\D 0^{\mu-1})^s(\D 0^{\mu \lambda}) \preccurlyeq R(n,S_C)^\rho.
\]
\begin{proof}
 See Lemma~\ref{lemma:LinAlgU}.
\end{proof}
\end{corollary}

 \begin{remark}\label{rem:D2}
  Proposition~\ref{boundD2U} is a generalization of the $\mathrm{BS}$ bound (Theorem~\ref{th:BSU}), and for the rare cases in which $\mu | n$, it is exactly the $\mathrm{BS}$ bound. 
 \end{remark}
\begin{remark}
 We note that bound II, when applicable, is sharper than bound I. In fact, if $(0^{\mu \lambda}\D)(0^{\mu-1}\D)^s \preccurlyeq R(n,S_C)^\rho$ for $\mu\geq2$, $s\geq \lambda+1$, in notation of Proposition~\ref{boundD1U} it means $(0^\ell \D^r)(0^{m}\D^r)^s \preccurlyeq R(n,S_C)^\rho$ with $\ell=\mu \lambda$, $r=1$, $m=\mu-1$ and then Proposition~\ref{boundD1U} gives a value $d_I$
 \[
d_I\geq \mu \lambda +1 +s-\left\lfloor\frac{\mu \lambda}{\mu}\right\rfloor-\max\Set{(\mu \lambda)_{\mu}-(\mu-1),0}=\mu \lambda+1+s-\lambda 
 \]
 while Proposition~\ref{boundD2U} gives a value $d_{II}$ 
 \[
d_{II}\geq\mu \lambda+\mu+s-\lambda-1  
 \]
and since $\mu\geq 2$ then $d_{II}\geq d_I$.
\end{remark}


\subsection{Proofs of bound I and bound II}
\label{proofs}

In this section we provide the proofs of Proposition~\ref{boundD1U}, and Proposition~\ref{boundD2U}.

\begin{remark}\label{remark:important}
The main tool we use to prove Proposition~\ref{boundD1U} is Theorem~\ref{th:Schaub} which, in principle, allows us to work only with matrices that have as entries just $0$ or $\Dp$. Nevertheless during the proof we use matrices that have also $\D$ as entry. A $\D$ can be either $0$ or $\Dp$, the correctness of the proof is not affected by either choice.     
\end{remark}
\begin{proof}[of Proposition~\ref{boundD1U}]
 The general plan of the proof is as follows. Thanks to Theorem~\ref{th:Schaub} we aim at proving that
\[
 \min\Set{\mathrm{prk}(M(\mathbf{v})) \mid \mathbf{v} \in \mathcal{A}(R(n,S_C))}\geq \ell +1 +s-r \left\lfloor\frac{\ell}{m+r}\right\rfloor-\max\Set{(\ell)_{m+r}-m,0}.
\]
In order to do that, for any $\mathbf{v}\in\mathcal{A}(n,S_C)$, we need to choose $\ell +s+1$ rows in $M(\vv)$ and we must prove that, discarding at most $ r \left\lfloor\frac{\ell}{m+r}\right\rfloor+\max\Set{(\ell)_{m+r}-m,0}$ rows, we actually obtain a set of rows for which the singleton procedure is successful. 

We can suppose w.l.o.g. that $i_{0}=n-\ell$ (see Lemma~\ref{lemma:LinAlgU}), so that:
\[
 \vv=\underbrace{\D\dots\D}_{r}(\underbrace{0 \dots 0}_{m}\underbrace{\D \dots \D}_{r})^{s}\dots \underbrace{0\dots 0\dots 0}_{\ell}.
\]

We introduce two notions releated to $\vv$ (\cite{CGC-cd-prep-maxbetti}). From now on, the meaning of $\vv$ is fixed.
\begin{definition}
 Let $1\leq i'\leq n$. We say that $i'$ is the {\bf primary pivot} of ${\bf v}$ if ${\bf v}[i']$ is
 the first $\Dp$ that occurs in $\vv$, i.e.
 $$
   i'= \min \{ h \mid \vv[h]=\Dp \} \,.
 $$ 
\end{definition}
We can suppose that $1\leq i' \leq r$, otherwise $\vv= 0^r (0^m \D^r)^s \dots 0^\ell$ and so $(0^{\ell+r+m}\D^r)(0^m\D^r)^{s-1}\preccurlyeq \vv$ (Definition~\ref{def:contained}) and the bound would be trivially satisfied, since it would give:
\begin{align*}
 d	&\geq \ell+r+m+1+s-1-\left\lfloor\frac{\ell+r+m}{m+r}\right\rfloor r-\max\Set{(\ell+m+r)_{m+r}-m,0}\\
	& = \ell+r+m+s-\left\lfloor\frac{\ell}{m+r}\right\rfloor r-\max\Set{(\ell)_{m+r}-m,0}\\
	&\geq\ell+r+1+s-\left\lfloor\frac{\ell}{m+r}\right\rfloor r-\max\Set{(\ell)_{m+r}-m,0}.
\end{align*}

\begin{definition}\label{lht2}
 Let $n, m , r, s \in \NN$ s. t. $m$, $s\geq1$, $n\geq m+r$ and $(n,m+r)\leq m$. 
$((0)^m(\D)^{r})^s \preccurlyeq {\bf v} $.
 Then there are $i''$ in $\{1, \dots, n\}$, $k \in \NN$ and 
 $t\in\{1,\dots ,m\}$, with the following properties: 
\begin{enumerate}
\item ${\bf v}[i''] = \Dp$,
\item $i'' \equiv (s+k)(m+r) + t \mod (n)$,
\item ${\bf v}[i]=0$, for any $i$ s.t. 
      $$i\equiv (s+k')(m+r) + j \mod(n)\,,$$ 
      where $k'\in\{0,\dots ,k-1\}$ and $j \in\{1,\dots, m\}$.
\end{enumerate}
We call such $i''$ the {\bf secondary pivot} of $\vv$ with respect to block $((0)^m(\D)^{r})^s$.
\end{definition}
The following lemma shows that if $(m+r,n)\leq m$ (which includes the classical case $(m+r,n)=1$), then the secondary pivot exists.\\

\begin{lemma}\label{lht}
  Let $n,m,r,s \in \NN$ such that $n\geq m+r$, $m \geq 1$, $s\geq 1$ and $(m+r,n)\leq m $. Then for any
  $i$ in $\{1,\dots,n\}$ there are $k \in \NN$ and $1 \leq t\leq m$ 
  such that
  \begin{equation}\label{eq:secondary}
   i \equiv (s+k)(m+r) + t  \mod (n).
  \end{equation}
\begin{proof}
 Given $i \in \{1,\dots,n\}$, let $\lambda =(m+r,n)$. 
 By hypothesis $\lambda\in\Set{0,\dots, m}$. 
 We take $t=(i)_{\lambda}$ and we note $t\in\Set{0,\dots, m-1}$. Let $\overline{k}\in \NN$ be such that $i-t \equiv \overline{k}\lambda$.
 Now, by Bézout's identity, there exist two integers $a$, $b$ s.t. $\lambda=a (m+r)+ b n$, so $i-t=	\overline{k}a(m+r)+ \overline{k}bn\implies i\equiv \overline{k}a(m+r)+t \mod (n)$.\\
 It is sufficient to take $k= (\overline{k}a -s)_{n}$ to satisfy the congruence \eqref{eq:secondary}.
\end{proof}
\end{lemma}

We can suppose $s(m+r)+r+1\leq i''\leq  s(m+r)+r+m$, otherwise we have $(0^{\ell}\D^r)(0^m\D^r)^{s+1}\preccurlyeq \vv$ and the bound is trivially satisfied:  
\begin{align*}
 d	&\geq \ell+1+s+1-\left\lfloor\frac{\ell}{m+r}\right\rfloor r-\max\Set{(\ell+m+r)_{m+r}-m,0}\\
	&\geq\ell+1+s-\left\lfloor\frac{\ell}{m+r}\right\rfloor r-\max\Set{(\ell)_{m+r}-m,0}.
\end{align*}

We note that $\vv[i''-z\cdot(m+r)]=0$ for any $z=1,\dots,s$. Moreover, $i'$ and $i''$ may coincide, but this is not a problem. 



Now, we are going to choose $(\ell+1+s)$ rows of $M(\vv)$. We start from 
the $((n-i'+k)_{n}+1)-$th rows with $k=1,\dots,m$, that is, we take the row with the primary pivot in the first position and its shifts up to the $(m-1)-$th shift included.
We collect these rows in submatrix $T_1$ and we note that they are clearly linearly independent, applying the singleton procedure (see \cite{CGC-cd-art-maxbetti}, Lemma~3.2).

$$
T_1=\left(
\begin{smallmatrix}
\Dp 	& \dots 	& 0 	& \dots 	& 0 	& \D 	& \dots 	& \D 	& \dots 	& 0 	& \dots 	& 0 	& \D 	& \dots 	& \Dp 	& \dots 	& \dots 	& \dots & 0 	& \dots 	& \dots	& 0     	\\
 0  	& \Dp 	& \dots 	&    0 	& \dots 	& 0 	& \D 	& \dots	& \D 	& \dots 	&  0 	& \dots 	&  0 	& \D 	& \dots 	& \Dp 	& \dots 	& \dots & \dots 	&  0 	& \dots 	& 0	\\
 0	&  0	&\Dp 	& \dots 	&    0 	& \dots 	& 0 	& \D 	& \dots	& \D 	& \dots 	&  0 	& \dots 	&  0 	& \D 	& \dots 	& \Dp 	& \dots & \dots 	& \dots	&   0 	& \dots	\\
\vdots  	&\vdots 	&\vdots 	&\vdots	&\vdots 	&\vdots	&\vdots 	&\vdots 	&\vdots	&\vdots 	&\vdots	&\vdots	&\vdots	&\vdots	&\vdots 	&\vdots 	&\vdots 	&\vdots &\vdots 	&\vdots 	&\vdots	&\vdots	\\
 0	& \dots 	& 0	& \Dp	&\dots	&    0 	& \dots 	& 0 	& \D 	& \dots	& \D 	& \dots 	&  0 	& \dots 	&  0 	& \D 	& \dots 	& \Dp 	& \dots & \dots 	& 0 	& \dots	\\ 
	&	&	&\downarrow&																		\\
	&	&	& m																			\\								
 \end{smallmatrix}
 \right)
$$
We now consider the $(k+1)$-th rows for $k=m,\dots, \ell$, collected in submatrix $T_2$.
$$
T_2=\left(
\begin{smallmatrix}
 0	& \dots 	& 0 	& \D 	& \dots 	& \D 	& \dots 	& 0 	& \dots 	& 0 	& \D 	& \dots 	& \D 	& \dots 	& \Dp 	& \dots 	& \dots 	& \dots & \dots 	& 0 	& \dots	& \dots	\\
 0	& \dots	& \dots	& 0 	& \D 	& \dots 	& \D 	& \dots 	& 0 	& \dots 	& 0 	& \D 	& \dots 	& \D 	& \dots 	& \Dp 	& \dots 	& \dots 	& \dots & \dots 	& 0 	& \dots	\\
 0	& \dots	& 0 	& \dots	& 0 	& \D 	& \dots 	& \D 	& \dots 	& 0 	& \dots 	& 0 	& \D 	& \dots 	& \D 	& \dots 	& \Dp 	& \dots 	& \dots 	& \dots & \dots 	& \dots	\\
\vdots  	&\vdots 	&\vdots 	&\vdots	&\vdots 	&\vdots	&\vdots 	&\vdots 	&\vdots	&\vdots 	&\vdots	&\vdots	&\vdots	&\vdots	&\vdots 	&\vdots 	&\vdots 	&\vdots &\vdots 	&\vdots 	&\vdots	&\vdots	\\
 0	&\dots  	& \dots	& 0	& \dots	& 0 	& \dots 	& 0 	& \D 	& \dots 	& \D 	& \dots 	& 0 	& \dots 	& 0 	& \D 	& \dots 	& \D 	& \dots 	& \Dp 	& \dots & \dots	\\
 0	&\dots  	& 0     	& \dots	& 0	& \dots	& 0 	& \dots 	& 0 	& \D 	& \dots 	& \D 	& \dots 	& 0 	& \dots 	& 0 	& \D 	& \dots 	& \D 	& \dots 	& \Dp 	& \dots 	\\ 
	&	&\downarrow&	&	&	&	&	&\downarrow&														\\																		
	&	& m	&	&	&	&	&	&\ell 															\\								
 \end{smallmatrix}
\right)
$$
Note that $T_1$ and $T_2$ have no common rows.
Note also that in $T_2$ for any row $h=1,\dots,\ell +1-m$ and any column $1\leq j \leq (s-1)(m+r)+m$ we have:  
\begin{equation}\label{eq:T2}
 T_2[h,j]=\D\implies T_2[h,j+(m+r)]=\D
\end{equation}
Moreover, $T_2$ has full rank as the following lemma shows.
\begin{lemma}\label{lem:subT2}
The singleton procedure is successful for $T_2$ and thus $\mathrm{prk}(T_2)=\ell-m+1$.
\begin{proof}
We are going to prove that the singleton procedure is successful for all the rows of $T_2$. We have that $\vv[i']=\Dp$ and $\vv[i]=0, \ \forall \ i \in \Set{i'-1,\dots, i'-\ell}$. In particular $\vv[i]=0, \ \forall \ i \in \Set{i'-1,\dots, i'-\ell+m}$.\\ We note that since every row of $T_2$ is obtained from a right-shift of the previous one and the first row of $T_2$ is obtained shifting $\vv$ of $m$ positions to the right, so for $1\leq h\leq \ell-m-2$ it holds
\begin{align*}
 T_2[h+1,j]=T_2[h,j-1]  && \mbox{and} && T_2[1,j]= \vv[j-m].
\end{align*}
At the first step we s-delete the first row and the $(i'+m)-$th column, since $T_{2}(i'+m)$ is a singleton, in fact for $2\leq h \leq \ell-m+1$:
\[
 T_2[h,i'+m]=T_2[1,i'+m-(h-1)]=\vv[i'-(h-1)]=0
\]
while $T_2[1,i'+m]=\vv[i']=\Dp$.\\
Suppose now we have s-deleted the first $j$ rows, we want to show that the matrix $T_2^{(j)}$ obtained from these $j$ s-deletions has a singleton in $T_2^{(j)}(i'+m+j)$. In fact, for $2\leq h\leq \ell-m+1-j$:
\begin{align*}
 T_2^{(j)}[h,(i'+m+j)]	& 	= T_2[j+h,(i'+m+j)] 	\\
			&	=T_2[1,i'+m-(h-1)] 	\\
			&	=\vv[i'-(h-1)]=0	
\end{align*}
while
$T_2^{(j)}[1,(i'+m+j)]= T_2[j+1,(i'+m+j)]=T_2[1,i'+m]=\vv[i']=\Dp$. After $(\ell-m)$ steps we have that $T_2^{(\ell-m)}$ is the last row of the matrix $T_2$, (i.e. $T_2^{(\ell-m)}=T_2[\ell-m+1]$), which is different from zero, since $T_2[\ell-m+1,i'+\ell+1]=T_2[1,i'+m]=\vv[i']=\Dp$. 
\end{proof}
\end{lemma}
Since all the rows of $T_2$ have a block of zeros in the first $m$-positions, they are linearly independent from all the rows in $T_1$. We can conclude that any matrix containing $T_1$ and $T_2$ has rank at least $\ell+1$, obtaining \eqref{eq1UUU}. 
If $(m+r,n)\leq m$ we can also consider a third and last submatrix, $T_3$, formed by the $((n-r-k\cdot(m+r))_n +1)-$th rows, for $k=0,\dots, (s-1)$:
$$
T_3=\left(
\begin{smallmatrix}
 0	& \dots 	& 0 	& \D 	& \dots 	& \D 	& \dots 	& 0	& \dots 	& 0 	& \D 	& \dots 	& \D 	& 0	& \dots 	& 0 	& \D 	& \dots 	& \D 	& \dots	& \Dp	& \dots	\\
 0	& \dots 	& 0 	& \D 	& \dots 	& \D 	& \dots 	& 0	& \dots 	& 0 	& \D 	& \dots 	& \D 	& \dots	& \Dp	& \dots	& \dots	& \dots	& \dots 	& \dots 	& \dots & \dots	\\
\vdots  	&\vdots 	&\vdots 	&\vdots	&\vdots 	&\vdots	&\vdots 	&\vdots 	&\vdots	&\vdots 	&\vdots	&\vdots	&\vdots	&\vdots	&\vdots 	&\vdots 	&\vdots 	&\vdots &\vdots 	&\vdots 	&\vdots	&\vdots	\\
 0	& \dots 	& 0 	& \D 	& \dots 	& \D 	& \dots	& \Dp	& \dots	& \dots	& \dots	& \dots 	& \dots 	& \dots & \dots	& \dots	& \dots	& \dots	& \dots 	& \dots 	& \dots & \dots	\\ 
	&	&\downarrow&	&	&\downarrow&	&\downarrow&	&	&	&	&	&	&	&	&	&	&	&	&\downarrow&	\\
	&	& m	&	&	& m+r	&	& i''-r-(s-1)(m+r)&	&	&	&	&	&	&	&	&	&	&	&	&i''-r		\\	
 \end{smallmatrix}
\right)
$$
\begin{lemma}\label{lem:subT3}
 The singleton procedure is successful for $T_3$ and thus $\mathrm{prk}(T_3)=s$. Moreover $T_3[k]$ is the row corresponding to the singleton $T_3(i''-r-(k-1)(m+r))$ for $k=s,s-1,\dots,1$.
\begin{proof}
We note that the rows of $T_3$, by construction, have the property that \\ $T_3[a+1,h]=T_3[a,h+(m+r)]$ because each row is a $(m+r)$ left shift of the previous one. This is sufficient to prove that $T_3(i'' -r-(s-1)(m+r))$ is a singleton. We claim that the $s-$th row of $T_3$ corresponds to a singleton. Indeed
{\small$$
T_3[s,i''-r-(s-1)(m+r)]=T_3[1,i''-r-(s-1)(m+r)+(s-1)(m+r)]=T_3[1,i''-r]=\Dp 
$$}
and for $k=1,\dots, s-1$:
{\small
$$
 T_3[k,i''-r-(s-1)(m+r)] = T_3[1,i''-r-(s-1)(m+r)+(k-1)(m+r)]= T_3[i''-r-(s-k)(m+r)]  =   0   	
$$
}
so we can s-delete it. Once this is done, we might also s-delete the $(s-1)-$th row, since
{\small$$
T_3[s-1,i''-r-(s-2)(m+r)]=T_3[1,i''-r-(s-2)(m+r)+(s-2)(m+r)]=T_3[1,i''-r]=\Dp 
$$}
and for $k=1,\dots, s-2$:
{\small
$$
 T_3[k,i''-r-(s-2)(m+r)] = T_3[i''-r-(s-2)(m+r)+(k-1)(m+r)]= T_3[1,i''-r-(s-1-k)(m+r)]  =   0.   	
$$
}
In this way for any row of $T_3$ we obtain a singleton in $T_3\left(i''-r-k(m+r)\right)$ for $k=0,\dots,s-1$, by recursively s-deleting from the last row to the first.
\end{proof}
\end{lemma}

Collecting all these submatrices $T_1$, $T_2$, $T_3$, we obtain an $(\ell+1+s) \times n$ matrix $T$, as follows:
\[
 T=\left(
\begin{smallmatrix}
\Dp 	& \dots 	& 0 	& \dots 	& 0 	& \D 	& \dots 	& \D 	& \dots 	& 0 	& \dots 	& 0 	& \D 	& \dots 	& \Dp 	& \dots 	& \dots 	& \dots & 0 	& \dots 	& \dots	& 0     	& \rightarrow	& 1			\\
0  	& \Dp 	& \dots 	&    0 	& \dots 	& 0 	& \D 	& \dots	& \D 	& \dots 	&  0 	& \dots 	&  0 	& \D 	& \dots 	& \Dp 	& \dots 	& \dots & \dots 	&  0 	& \dots 	& 0	&		&			\\
\vdots  	&\vdots 	&\vdots 	&\vdots	&\vdots 	&\vdots	&\vdots 	&\vdots 	&\vdots	&\vdots 	&\vdots	&\vdots	&\vdots	&\vdots	&\vdots 	&\vdots 	&\vdots 	&\vdots &\vdots 	&\vdots 	&\vdots	& \vdots	&		&	& {\bf T_1}	\\
0	&  0	&\Dp 	& \dots 	&    0 	& \dots 	& 0 	& \D 	& \dots	& \D 	& \dots 	&  0 	& \dots 	&  0 	& \D 	& \dots 	& \Dp 	& \dots & \dots 	& \dots	&   0 	& \dots	&		&			\\
\\
\hline
\\
0	& \dots 	& 0 	& \D 	& \dots 	& \D 	& \dots 	& 0 	& \dots 	& 0 	& \D 	& \dots 	& \D 	& \dots 	& \Dp 	& \dots 	& \dots 	& \dots & \dots 	& 0 	& \dots	& \dots	& \rightarrow	& m+1		\\
0	& \dots	& \dots	& 0 	& \D 	& \dots 	& \D 	& \dots 	& 0 	& \dots 	& 0 	& \D 	& \dots 	& \D 	& \dots 	& \Dp 	& \dots 	& \dots 	& \dots & \dots 	& 0 	& \dots					\\
0	& \dots	& 0 	& \dots	& 0 	& \D 	& \dots 	& \D 	& \dots 	& 0 	& \dots 	& 0 	& \D 	& \dots 	& \D 	& \dots 	& \Dp 	& \dots 	& \dots 	& \dots 	& \dots 	& \dots					\\
\vdots  	&\vdots 	&\vdots	&\vdots 	&\vdots	&\vdots 	&\vdots 	&\vdots	&\vdots 	&\vdots	&\vdots	&\vdots	&\vdots	&\vdots 	&\vdots 	&\vdots 	&\vdots &\vdots 	&\vdots 	&\vdots	&\vdots	& \dots	&		&	& {\bf T_2}	\\
0	&\dots  	& 0	& \dots	& 0 	& \dots 	& 0 	& \D 	& \dots 	& \D 	& \dots 	& 0 	& \dots 	& 0 	& \D 	& \dots 	& \D 	& \dots 	& \Dp 	& \dots 	& \dots	& \dots					\\
0	&\dots  	& 0     	& \dots	& 0	& \dots	& 0 	& \dots 	& 0 	& \D 	& \dots 	& \D 	& \dots 	& 0 	& \dots 	& 0 	& \D 	& \dots 	& \D 	& \dots 	& \Dp 	& \dots	& \rightarrow	& \ell+1	\\
\\
\hline
\\
0	& \dots 	& 0 	& \D 	& \dots 	& \D 	& \dots 	& 0	& \dots 	& 0 	& \D 	& \dots 	& \D 	& 0	& \dots 	& 0 	& \D 	& \dots 	& \D 	& \dots	& \Dp	& \dots	\\
0	& \dots 	& 0 	& \D 	& \dots 	& \D 	& \dots 	& 0	& \dots 	& 0 	& \D 	& \dots 	& \D 	& \dots	& \Dp	& \dots	& \dots	& \dots	& \dots 	& \dots 	& \dots & \dots	\\
\vdots  	&\vdots 	&\vdots 	&\vdots	&\vdots 	&\vdots	&\vdots 	&\vdots 	&\vdots	&\vdots 	&\vdots	&\vdots	&\vdots	&\vdots	&\vdots 	&\vdots 	&\vdots 	&\vdots &\vdots 	&\vdots 	&\vdots	& \dots	&		&	& {\bf T_3}	\\
0	& \dots 	& 0 	& \D 	& \dots 	& \D 	& \dots	& \Dp	& \dots	& \dots	& \dots	& \dots 	& \dots 	& \dots & \dots	& \dots	& \dots	& \dots	& \dots 	& \dots 	& \dots & \dots	& \rightarrow	& \ell+1+s	\\ 
\\
	&	&\downarrow&	& 	&\downarrow&	\\
	&	& m	&	&	& m+r	&	\\
\end{smallmatrix}
 \right)
\]
Observe that the rows from $(m+1)$ to $(\ell +s+1)$ have a block of zero in the first $m$ positions, so we can obviously s-delete the first $m$ rows (i.e the rows of $T_1$). After these first $m$ s-deletions we obtain a matrix $T'$ composed of the last $(\ell+1+s-m)$ rows of $T$, as the following:
\[
 T'=\left(
\begin{smallmatrix}
0	& \dots 	& 0 	& \D 	& \dots 	& \D 	& \dots 	& 0 	& \dots 	& 0 	& \D 	& \dots 	& \D 	& \dots 	& \Dp 	& \dots 	& \dots 	& \dots & \dots 	& 0 	& \dots	& \dots	& \rightarrow	& m+1	\\
0	& \dots	& \dots	& 0 	& \D 	& \dots 	& \D 	& \dots 	& 0 	& \dots 	& 0 	& \D 	& \dots 	& \D 	& \dots 	& \Dp 	& \dots 	& \dots 	& \dots & \dots 	& 0 	& \dots	\\
0	& \dots	& 0 	& \dots	& 0 	& \D 	& \dots 	& \D 	& \dots 	& 0 	& \dots 	& 0 	& \D 	& \dots 	& \D 	& \dots 	& \Dp 	& \dots 	& \dots 	& \dots 	& \dots 	& \dots	\\
\vdots  	&\vdots 	&\vdots	&\vdots 	&\vdots	&\vdots 	&\vdots 	&\vdots	&\vdots 	&\vdots	&\vdots	&\vdots	&\vdots	&\vdots 	&\vdots 	&\vdots 	&\vdots &\vdots 	&\vdots 	&\vdots	&\vdots	& \dots	\\
0	&\dots  	& 0	& \dots	& 0 	& \dots 	& 0 	& \D 	& \dots 	& \D 	& \dots 	& 0 	& \dots 	& 0 	& \D 	& \dots 	& \D 	& \dots 	& \Dp 	& \dots 	& \dots	& \dots	\\
0	&\dots  	& 0     	& \dots	& 0	& \dots	& 0 	& \dots 	& 0 	& \D 	& \dots 	& \D 	& \dots 	& 0 	& \dots 	& 0 	& \D 	& \dots 	& \D 	& \dots 	& \Dp 	& \dots	& \rightarrow	& \ell+1	\\
\\
0	& \dots 	& 0 	& \D 	& \dots 	& \D 	& \dots 	& 0	& \dots 	& 0 	& \D 	& \dots 	& \D 	& 0	& \dots 	& 0 	& \D 	& \dots 	& \D 	& \dots	& \Dp	& \dots	\\
0	& \dots 	& 0 	& \D 	& \dots 	& \D 	& \dots 	& 0	& \dots 	& 0 	& \D 	& \dots 	& \D 	& \dots	& \Dp	& \dots	& \dots	& \dots	& \dots 	& \dots 	& \dots & \dots	\\
\vdots  	&\vdots 	&\vdots 	&\vdots	&\vdots 	&\vdots	&\vdots 	&\vdots 	&\vdots	&\vdots 	&\vdots	&\vdots	&\vdots	&\vdots	&\vdots 	&\vdots 	&\vdots 	&\vdots &\vdots 	&\vdots 	&\vdots	& \dots	\\
0	& \dots 	& 0 	& \D 	& \dots 	& \D 	& \dots	& \Dp	& \dots	& \dots	& \dots	& \dots 	& \dots 	& \dots & \dots	& \dots	& \dots	& \dots	& \dots 	& \dots 	& \dots & \dots	& \rightarrow	& \ell+1+s	\\ 
\\
	&	&\downarrow&	& 	&\downarrow&	&	&	&	&	&	&	&	&	&	&	&	&\downarrow&	&\downarrow&	\\
	&	& m	&	&	& m+r	&	&	&	&	&	&	&	&	&	&	&	&	& s(m+r)	&	& i''-r	&	\\
\end{smallmatrix}
 \right)
\]
where $1+s(m+r)\leq i''-r\leq m+ s(m+r)$ by hypothesis. We note that $T'$ is composed by the rows of $T_2$ and $T_3$.\\

We use the singletons of $T_3$ to proceed with the singleton procedure, but in order to do that we have to discard some rows in $T_2$. More precisely, let us define:
\begin{align*}
 B_{k}=\Set{h \mid T_2[h,i''-r-k(m+r)]=\D} && \text{ for } k=0,\dots, s-1
\end{align*}
then the rows to discard in $T_2$ in order that $T(i''-r-k(m+r))$ becomes a singleton for $k=0,\dots, s-1$ are: 
\begin{equation}\label{eq:BB}
\mathbf{B}=	\cup_{k=0}^{s-1}B_{k}.
\end{equation}
\begin{lemma}
 Let $0 \leq k<k'\leq s-1 $, then $B_{k'}\subseteq B_{k}$.
 \begin{proof}
  Obvious from (\ref{eq:T2}).
 \end{proof}
\end{lemma}
\begin{corollary}\label{cor:B}
 $\mathbf{B}=B_{0}=\Set{h \mid T_2[h,i''-r]=\D}$.
\end{corollary}
Thanks to Corollary \ref{cor:B}, since $s(m+r)+1\leq i''-r \leq s(m+r)+m$, if we define $\eta_j= | \Set{h \mid T_2[h,s(m+r)+j]=\D} |$, we have: 
\[
|\mathbf{B}|\leq \max\Set{ \eta_j \mid 1\leq j \leq m} .
\]
and we can further improve this result with the following lemma, which is not difficult to prove.
\begin{lemma}\label{lem:eta}
 For $1\leq j\leq m$:
 \[
  \eta_{1}\geq \eta_{2} \geq \dots \geq \eta_{m}.
 \]
\end{lemma}
Thanks to lemma \ref{lem:eta} we are able to estimate the maximal number of rows of $T_2$ that we have to discard.
\begin{lemma}\label{lem:BB}
 \[
  |\mathbf{B}|\leq \eta_{1}\leq \left\lfloor\frac{\ell}{m+r}\right\rfloor r + \max\Set{(\ell)_{m+r}-m,0}
 \]
\begin{proof}
 For Corollary \ref{cor:B} and Lemma \ref{lem:eta} we have $|\mathbf{B}|\leq \eta_{1}$. Now:
 $$
  \eta_{1}= | \Set{h \mid T_2[h,s(m+r)+1]=\D} |, \text{ but recall } 1\leq h \leq \ell+1-m.
  $$
  We rewrite $\vv$ in the worst case where $i''=s(m+r)+r+1$:
\[       
 \begin{smallmatrix}
  \vv=
  &\D 		& \dots & \D 		& 0 	& \dots 	& 0 		& (\D^r 0^m)^{s-2}	&	\D	&	\dots 	&	\D 	& 0 \ \	\ \ \ \ \ \ \ \dots	& 	& 0 		& \D 	& \dots 	& \D 	&\Dp	&\dots \ \dots	&	\\
  &\downarrow	& 	&\downarrow	& 	& 	& \downarrow	& 			&		&		&		&\downarrow			& 	&\downarrow 	& 	& 	& 	& \downarrow	& 	\\
  &1		& 	& r		& 	& 	& m+r		& 			&	 	& 		& 		&s(m+r)-m+1			& 	&s(m+r)		& 	& 	& 	& s(m+r)+r+1	& 	\\
 \end{smallmatrix}
\]
Since $T_2[1,s(m+r)+1]=\vv[s(m+r)+1-m]=0$, we have
  \begin{align*}
  \eta_{1}	&	= | \Set{h \mid T_2[h,s(m+r)+1]=\D, 1\leq h \leq \ell+1-m} |		\\
		&	= | \Set{h \mid T_2[h,s(m+r)+1]=\D, 2\leq h \leq \ell+1-m} |. 	
  \end{align*}
Now $ T_2[h+1,j]=T_2[h,j-1]$ (for $h\geq 1$) and $T_2[1,j]= \vv[j-m]$, by construction of $T_2$. So:
 \begin{align*}
  \eta_{1} 	&	= | \Set{h \mid T_2[h,s(m+r)+1]=\D, 2\leq h \leq \ell+1-m} | 		\\
		&	= | \Set{h \mid T_2[1,s(m+r)+1-(h-1)]=\D, 2\leq h \leq \ell+1-m} |		\\
		&	= | \Set{h \mid \vv[s(m+r)-m+2-h]=\D, 2\leq h \leq \ell+1-m} |		\\
		&	= | \Set{h \mid \vv[s(m+r)+2-h]=\D, 2\leq h \leq \ell+1} |		
 \end{align*}
Thus, to compute $\eta_1$ we have to count the number of $\D$'s we encounter, from $\vv[s(m+r)]$ to $\vv[s(m+r)-\ell+1]$ (i.e. from $\vv[s(m+r)]$ and going back $\ell$ positions). Let us consider the worst case, which is when $\ell\leq s(m+r)$. Passing through the block $(0^m\D^r)^s$ from right to left through $\ell$ positions, every $m+r$ steps we meet a block formed by $r$ $\D$'s and $m$ $0$'s, thus the contribution to $\eta_{1}$ per block is $r$. Since we move by $\ell$ positions only, we cross no more than $\left\lfloor\frac{\ell}{m+r}\right\rfloor$ such blocks and so we have $\eta_{1}\leq \left\lfloor\frac{\ell}{m+r}\right\rfloor r+\eta_{1}'$, where $\eta_{1}'$ are the $\D$'s coming from the last $(\ell)_{m+r}$ steps left. The first $m$-positions we meet doing the last $(\ell)_{m+r}$ steps are zero, since they correspond to the last block $(\D^r 0^m)$, thus $\eta_1'$ can be at most $(\ell)_{m+r}-m$ and it is non-negative only if $(\ell)_{m+r}\geq m$.
In conclusion: $\eta_{1}\leq \left\lfloor\frac{\ell}{m+r}\right\rfloor r + \max\Set{(\ell)_{m+r}-m,0}$.
\end{proof}
\end{lemma}

Thanks to Lemma~\ref{lem:BB}, discarding at most $\left\lfloor\frac{\ell}{m+r}\right\rfloor r + \max\Set{(\ell)_{m+r}-m,0}$ rows of $T_2$, we can remove by s-deletions $T_3$ from $T'$.The matrix that remains is a submatrix $\widetilde{T}$ of $T_2$ not having row indeces in $\mathbf{B}$. Note that $\widetilde{T}$ has full rank, because $T_2$ has full rank by Lemma~\ref{lem:subT2}. So we have proved Proposition~\ref{boundD1U}. \qed  
\end{proof}

\begin{example}
 Let $C$ be a cyclic code of length $n$, with defining set $S_C$ satisfying the assumptions of Proposition~\ref{boundD1U} with parameters $\ell=7$, $m=2$, $r=1$, $s=5$. We want to prove that by Proposition~\ref{boundD1U} the distance of the code $C$ is at least $d\geq 7+1+5-\left\lfloor\frac{7}{2+1}\right\rfloor 1 -\max\Set{\left(7\right)_{3+2}-2,0}=11.$ 
 Let $\vv\in\mathcal{A}(R(n,S_C))$ with $\vv[1]=\Dp$. The matrix $T$ is:
 $$\left(
\begin{smallmatrix}
  \Dp 	& 0	& 0 	& \D 	& 0 	& 0 	& \D 	& 0 	& 0 	& \D 	& 0 	& 0 	& \D 	& 0 	& 0 	& \D 	& \Dp 	& \D  	& \D  	& \D  	& \D  	& \D  	& \D  	& \D  	& \D  	& \dots 	& \dots \\
   0 	& \Dp 	& 0  	& 0 	& \D 	& 0 	& 0 	& \D 	& 0 	& 0 	& \D 	& 0 	& 0 	& \D 	& 0 	& 0 	& \D 	& \Dp 	& \D  	& \D  	& \D  	& \D  	& \D  	& \D  	& \D  	& \dots 	& \dots \\
\\
\hline
\\
   0 	& 0 	& \Dp	& 0 	& 0 	& \D 	& 0 	& 0 	& \D 	& 0 	& 0 	& \D 	& 0 	& 0 	& \D 	& 0 	& 0 	& \D 	& \Dp 	& \D  	& \D  	& \D  	& \D  	& \D  	& \D  	& \dots 	& \dots \\
   0 	& 0 	& 0  	& \Dp 	& 0 	& 0 	& \D 	& 0 	& 0 	& \D 	& 0 	& 0 	& \D 	& 0 	& 0 	& \D 	& 0 	& 0 	& \D 	& \Dp 	& \D  	& \D  	& \D  	& \D  	& \D  	& \dots 	& \dots \\
   0 	& 0 	& 0  	& 0 	& \Dp 	& 0 	& 0 	& \D 	& 0 	& 0 	& \D 	& 0 	& 0 	& \D 	& 0 	& 0 	& \D 	& 0 	& 0 	& \D 	& \Dp 	& \D  	& \D  	& \D  	& \D  	& \dots 	& \dots \\
   0 	& 0 	& 0  	& 0 	& 0 	& \Dp 	& 0 	& 0 	& \D 	& 0 	& 0 	& \D 	& 0 	& 0 	& \D 	& 0 	& 0 	& \D 	& 0 	& 0 	& \D 	& \Dp 	& \D  	& \D  	& \D  	& \dots 	& \dots \\
   0 	& 0 	& 0 	& 0 	& 0 	& 0 	& \Dp 	& 0 	& 0 	& \D 	& 0 	& 0 	& \D 	& 0 	& 0 	& \D 	& 0 	& 0 	& \D 	& 0 	& 0 	& \D 	& \Dp 	& \D  	& \D  	& \dots 	& \dots \\
   0 	& 0 	& 0 	& 0 	& 0 	& 0 	& 0 	& \Dp 	& 0 	& 0 	& \D 	& 0 	& 0 	& \D 	& 0 	& 0 	& \D 	& 0 	& 0 	& \D 	& 0 	& 0 	& \D 	& \Dp 	& \D  	& \dots 	& \dots \\ 
\\
\hline
\\
   0 	& 0 	& \D 	& 0 	& 0 	& \D 	& 0 	& 0 	& \D 	& 0 	& 0 	& \D 	& 0 	& 0 	& \D 	& \Dp 	& \D  	& \D  	& \D  	& \D  	& \D  	& \D  	& \D 	& \D 	& \D  	& \dots  	& \dots	\\
   0 	& 0 	& \D 	& 0 	& 0 	& \D 	& 0 	& 0 	& \D 	& 0 	& 0 	& \D 	& \Dp 	& \D  	& \D  	& \D  	& \D  	& \D  	& \D  	& \D 	& \D 	& \D  	& \D  	& \D	& \D  	& \dots  	& \dots	\\
   0 	& 0 	& \D 	& 0 	& 0 	& \D 	& 0 	& 0 	& \D 	& \Dp 	& \D  	& \D  	& \D  	& \D  	& \D  	& \D  	& \D 	& \D 	& \D  	& \D  	& \D	& \D  	& \D  	& \D	& \D  	& \dots  	& \dots	\\
   0 	& 0 	& \D 	& 0 	& 0 	& \D 	& \Dp 	& \D  	& \D  	& \D  	& \D  	& \D  	& \D  	& \D 	& \D 	& \D  	& \D  	& \D	& \D  	& \D  	& \D	& \D  	& \D  	& \D	& \D  	& \dots  	& \dots	\\
   0 	& 0 	& \D 	& \Dp 	& \D  	& \D  	& \D  	& \D  	& \D  	& \D  	& \D 	& \D 	& \D  	& \D  	& \D	& \D  	& \D  	& \D	& \D  	& \D  	& \D	& \D  	& \D  	& \D	& \D	& \dots	& \dots	\\
\end{smallmatrix}\right)
$$
For the secondary pivot we have two possibilities: $i''=11$ or $i''=12$. We show that in both cases it is possible to obtain $11$ s-deletions, removing at most $\left\lfloor\frac{7}{2+1}\right\rfloor 1 +\max\Set{\left(7\right)_{3+2}-2,0}=2$ rows from the matrix $T$. \\

\emph{ Case 1: $i''=11$}. 
 $$
\begin{smallmatrix}
 \Dp 	& 0	& 0 	& \D 	& 0 	& 0 	& \D 	& 0 	& 0 	& \D 	& 0 	& 0 	& \D 	& 0 	& 0 	& \D 	& \Dp 	& \D  	& \D  	& \D  	& \D  	& \D  	& \D  	& \D  	& \D  	& \dots 	& \rightarrow & \mbox{\scriptsize{$1$-st s-deletion}} 	\\
   0 	& \Dp	& 0  	& 0 	& \D 	& 0 	& 0 	& \D 	& 0 	& 0 	& \D 	& 0 	& 0 	& \D 	& 0 	& 0 	& \D 	& \Dp 	& \D  	& \D  	& \D  	& \D  	& \D  	& \D  	& \D  	& \dots 	& \rightarrow & \mbox{\scriptsize{$2$-nd s-deletion}} 	\\  
   0	& 0	& \Dp	& 0 	& 0 	& \D 	& 0 	& 0 	& \D 	& 0 	& 0 	& \D 	& 0 	& 0 	& \D 	& 0 	& 0 	& \D 	& \Dp 	& \D  	& \D  	& \D  	& \D  	& \D  	& \D  	& \dots 	& \rightarrow & \mbox{\scriptsize{$8$-th s-deletion}} 	\\
\rlap{\rule[1mm]{0.8\textwidth}{0.5mm}}
   0	& 0	& 0  	& \Dp 	& 0 	& 0 	& \D 	& 0 	& 0 	& \D 	& 0 	& 0 	& \D 	& 0 	& 0 	& \D 	& 0 	& 0 	& \D 	& \Dp 	& \D  	& \D  	& \D  	& \D  	& \D  	& \dots 	& \rightarrow & \mathbf{REMOVED}		\\		
   0	& 0	& 0  	& 0 	& \Dp 	& 0 	& 0 	& \D 	& 0 	& 0 	& \D 	& 0 	& 0 	& \D 	& 0 	& 0 	& \D 	& 0 	& 0 	& \D 	& \Dp 	& \D  	& \D  	& \D  	& \D  	& \dots 	& \rightarrow & \mbox{\scriptsize{$9$-th s-deletion}} 	\\
   0	& 0	& 0  	& 0 	& 0 	& \Dp 	& 0 	& 0 	& \D 	& 0 	& 0 	& \D 	& 0 	& 0 	& \D 	& 0 	& 0 	& \D 	& 0 	& 0 	& \D 	& \Dp 	& \D  	& \D  	& \D  	& \dots 	& \rightarrow & \mbox{\scriptsize{$10$-th s-deletion}} 	\\
\rlap{\rule[1mm]{0.8\textwidth}{0.5mm}}
   0	& 0	& 0 	& 0 	& 0 	& 0 	& \Dp 	& 0 	& 0 	& \D 	& 0 	& 0 	& \D 	& 0 	& 0 	& \D 	& 0 	& 0 	& \D 	& 0 	& 0 	& \D 	& \Dp 	& \D  	& \D  	& \dots 	& \rightarrow & \mathbf{REMOVED}		\\
   0	& 0	& 0 	& 0 	& 0 	& 0 	& 0 	& \Dp 	& 0 	& 0 	& \D 	& 0 	& 0 	& \D 	& 0 	& 0 	& \D 	& 0 	& 0 	& \D 	& 0 	& 0 	& \D 	& \Dp 	& \D  	& \dots 	& \rightarrow & \mbox{\scriptsize{$11$-th s-deletion}} 	\\
   0	& 0	& \D 	& 0 	& 0 	& \D 	& 0 	& 0 	& \D 	& 0 	& 0 	& \D 	& 0 	& 0 	& \D 	& \Dp 	& \D  	& \D  	& \D  	& \D  	& \D  	& \D  	& \D 	& \D 	& \D  	& \dots 	& \rightarrow & \mbox{\scriptsize{$7$-th s-deletion}} 	\\
   0	& 0	& \D 	& 0 	& 0 	& \D 	& 0 	& 0 	& \D 	& 0 	& 0 	& \D 	& \Dp 	& \D  	& \D  	& \D  	& \D  	& \D  	& \D  	& \D 	& \D 	& \D  	& \D  	& \D	& \D  	& \dots 	& \rightarrow & \mbox{\scriptsize{$6$-th s-deletion}} 	\\
   0	& 0	& \D 	& 0 	& 0 	& \D 	& 0 	& 0 	& \D 	& \Dp 	& \D  	& \D  	& \D  	& \D  	& \D  	& \D  	& \D 	& \D 	& \D  	& \D  	& \D	& \D  	& \D  	& \D	& \D  	& \dots 	& \rightarrow & \mbox{\scriptsize{$5$-th s-deletion}} 	\\
   0	& 0	& \D 	& 0 	& 0 	& \D 	& \Dp 	& \D  	& \D  	& \D  	& \D  	& \D  	& \D  	& \D 	& \D 	& \D  	& \D  	& \D	& \D  	& \D  	& \D	& \D  	& \D  	& \D	& \D  	& \dots 	& \rightarrow & \mbox{\scriptsize{$4$-th s-deletion}} 	\\ 
   0	& 0	& \D 	& \Dp 	& \D  	& \D  	& \D  	& \D  	& \D  	& \D  	& \D 	& \D 	& \D  	& \D  	& \D	& \D  	& \D  	& \D	& \D  	& \D  	& \D	& \D  	& \D  	& \D	& \D	& \dots	& \rightarrow & \mbox{\scriptsize{$3$-rd s-deletion}} 	\\
\end{smallmatrix}
$$

\emph{ Case 2: $i''=12$}. \\
 $$
\begin{smallmatrix}
 \Dp 	& 0	& 0 	& \D 	& 0 	& 0 	& \D 	& 0 	& 0 	& \D 	& 0 	& 0 	& \D 	& 0 	& 0 	& \D 	& \D 	& \Dp  	& \D  	& \D  	& \D  	& \D  	& \D  	& \D  	& \D  	& \dots 	& \rightarrow & \mbox{\scriptsize{$1$-st s-deletion}} 	\\
   0 	& \Dp	& 0  	& 0 	& \D 	& 0 	& 0 	& \D 	& 0 	& 0 	& \D 	& 0 	& 0 	& \D 	& 0 	& 0 	& \D 	& \D 	& \Dp  	& \D  	& \D  	& \D  	& \D  	& \D  	& \D  	& \dots 	& \rightarrow & \mbox{\scriptsize{$2$-nd s-deletion}} 	\\  
   0	& 0	& \Dp	& 0 	& 0 	& \D 	& 0 	& 0 	& \D 	& 0 	& 0 	& \D 	& 0 	& 0 	& \D 	& 0 	& 0 	& \D 	& \D 	& \Dp  	& \D  	& \D  	& \D  	& \D  	& \D  	& \dots 	& \rightarrow & \mbox{\scriptsize{$8$-th s-deletion}} 	\\
   0	& 0	& 0  	& \Dp 	& 0 	& 0 	& \D 	& 0 	& 0 	& \D 	& 0 	& 0 	& \D 	& 0 	& 0 	& \D 	& 0 	& 0 	& \D 	& \D 	& \Dp  	& \D  	& \D  	& \D  	& \D  	& \dots 	& \rightarrow & \mbox{\scriptsize{$9$-th s-deletion}} 	\\
   \rlap{\rule[0.8mm]{0.8\textwidth}{0.4mm}}
   0	& 0	& 0  	& 0 	& \Dp 	& 0 	& 0 	& \D 	& 0 	& 0 	& \D 	& 0 	& 0 	& \D 	& 0 	& 0 	& \D 	& 0 	& 0 	& \D 	& \D 	& \Dp  	& \D  	& \D  	& \D  	& \dots 	& \rightarrow & \mathbf{REMOVED}		\\
   0	& 0	& 0  	& 0 	& 0 	& \Dp 	& 0 	& 0 	& \D 	& 0 	& 0 	& \D 	& 0 	& 0 	& \D 	& 0 	& 0 	& \D 	& 0 	& 0 	& \D 	& \D 	& \Dp  	& \D  	& \D  	& \dots 	& \rightarrow & \mbox{\scriptsize{$10$-th s-deletion}} 	\\
   0	& 0	& 0 	& 0 	& 0 	& 0 	& \Dp 	& 0 	& 0 	& \D 	& 0 	& 0 	& \D 	& 0 	& 0 	& \D 	& 0 	& 0 	& \D 	& 0 	& 0 	& \D 	& \D 	& \Dp  	& \D  	& \dots 	& \rightarrow & \mbox{\scriptsize{$11$-th s-deletion}} 	\\
   \rlap{\rule[0.8mm]{0.8\textwidth}{0.4mm}}
   0	& 0	& 0 	& 0 	& 0 	& 0 	& 0 	& \Dp 	& 0 	& 0 	& \D 	& 0 	& 0 	& \D 	& 0 	& 0 	& \D 	& 0 	& 0 	& \D 	& 0 	& 0 	& \D 	& \D 	& \Dp  	& \dots 	& \rightarrow & \mathbf{REMOVED}		\\
   0	& 0	& \D 	& 0 	& 0 	& \D 	& 0 	& 0 	& \D 	& 0 	& 0 	& \D 	& 0 	& 0 	& \D 	& \D 	& \Dp  	& \D  	& \D  	& \D  	& \D  	& \D  	& \D 	& \D 	& \D  	& \dots 	& \rightarrow & \mbox{\scriptsize{$7$-th s-deletion}} 	\\
   0	& 0	& \D 	& 0 	& 0 	& \D 	& 0 	& 0 	& \D 	& 0 	& 0 	& \D 	& \D 	& \Dp  	& \D  	& \D  	& \D  	& \D  	& \D  	& \D 	& \D 	& \D  	& \D  	& \D	& \D  	& \dots 	& \rightarrow & \mbox{\scriptsize{$6$-th s-deletion}} 	\\
   0	& 0	& \D 	& 0 	& 0 	& \D 	& 0 	& 0 	& \D 	& \D 	& \Dp  	& \D  	& \D  	& \D  	& \D  	& \D  	& \D 	& \D 	& \D  	& \D  	& \D	& \D  	& \D  	& \D	& \D  	& \dots 	& \rightarrow & \mbox{\scriptsize{$5$-th s-deletion}} 	\\
   0	& 0	& \D 	& 0 	& 0 	& \D 	& \D 	& \Dp  	& \D  	& \D  	& \D  	& \D  	& \D  	& \D 	& \D 	& \D  	& \D  	& \D	& \D  	& \D  	& \D	& \D  	& \D  	& \D	& \D  	& \dots 	& \rightarrow & \mbox{\scriptsize{$4$-th s-deletion}} 	\\ 
   0	& 0	& \D 	& \D 	& \Dp  	& \D  	& \D  	& \D  	& \D  	& \D  	& \D 	& \D 	& \D  	& \D  	& \D	& \D  	& \D  	& \D	& \D  	& \D  	& \D	& \D  	& \D  	& \D	& \D	& \dots	& \rightarrow & \mbox{\scriptsize{$3$-th s-deletion}} 	\\
\end{smallmatrix}
$$
\end{example}
In a similar way we prove Proposition~\ref{boundD2U}.
\begin{proof}[of Proposition~\ref{boundD2U}]
We can suppose w.l.o.g. that $i_{0}=n-\lambda\mu$ (see Lemma~\ref{lemma:LinAlgU}), so that $\vv=\D\overbrace{\underbrace{0\dots 0\D}_{\mu}\dots\dots\underbrace{0\dots 0\D}_{\mu}}^{s-\text{times}}\dots\underbrace{0\dots 0}_{\mu \lambda}$. Let $i'$  and $i''$ be respectively the primary pivot and the secondary pivot of $\vv$. 
We can consider a simpler situation, that is, $i'=1$ and $s\mu+2\leq i''\leq  s\mu+\mu$. In fact, if $i'\neq 1$, then $(0^{(\lambda+1)\mu}\D)(0^\lambda\D)^{s-1}\preccurlyeq \vv$ and we have two cases:
 \begin{enumerate}[i)]
  \item if $s\geq \lambda+3$ then $s-1\geq \lambda+2$ and the bound would be satisfied since it holds:
  \[
   d\geq (\lambda+1)\mu+\mu+s-1-\lambda-2\geq (\lambda+1)\mu+s-\lambda-1\geq \lambda\mu+\mu+s-\lambda-1;
  \]

  \item if $s=\lambda+1, \lambda+2$ then $1\geq s-(\lambda+1)$ and so from the $\BCH$ bound we have:
  \[
   d\geq \lambda\mu+\mu+1\geq (\lambda+1)\mu+s-\lambda-1=(\lambda+1)\mu+s-(\lambda+1).
  \]
 \end{enumerate}
As regards $s\mu+2\leq i''\leq  s\mu+\mu$, if it does not hold we have $(0^{\mu \lambda}\D)(0^{\mu-1}\D)^{s+1}\preccurlyeq \vv$ and   
\begin{align*}
 d	&\geq \mu \lambda+\mu+s+1-\lambda-1\\
	&\geq\mu \lambda+\mu+s-\lambda-1.
\end{align*}

In a similar way to the proof of Proposition~\ref{boundD1U} we are going to choose $\lambda\mu+\mu+s$ rows of $M(\vv)$. 
We collect the first $(\lambda\mu+\mu)$ rows of $M(\vv)$ in a matrix $T_1$, noting that they are the row with the primary pivot in first position and its shifts up to the $(\lambda\mu+\mu-1)-$th shift (included), so:
$$
T_1=\left(
\begin{smallmatrix}
\Dp 	&  0 	& \dots 	& 0 	& \D 	& 0 	& \dots 	& 0 	& \D 	& 0 	& \dots 	& 0 	& \D 	& \dots 	& \Dp 	& \dots 	& \dots & 0 	& \dots 	& \dots	&\dots 	& 0     	& \rightarrow &	1 		\\
 0  	& \Dp 	& 0 	& \dots 	& 0 	& \D 	& 0 	& \dots 	& 0 	& \D 	& 0 	& \dots 	& 0 	& \D 	& \dots 	& \Dp 	& \dots 	& \dots & 0 	& \dots 	& \dots	&\dots 			 		\\
 0	&  0	& \Dp 	& 0 	& \dots 	& 0 	& \D 	& 0 	& \dots 	& 0 	& \D 	& 0 	& \dots 	& 0 	& \D 	& \dots 	& \Dp 	& \dots 	& \dots & 0 	& \dots 	&\dots 					\\
 0 	&\dots	& 0	& \Dp 	& 0 	& \dots 	& 0 	& \D 	& 0 	& \dots 	& 0 	& \D 	& 0 	& \dots 	& 0 	& \D 	& \dots 	& \Dp 	& \dots 	& \dots 	&\dots 	&\dots					\\
\vdots  	&\vdots 	&\vdots 	&\vdots	&\vdots 	&\vdots	&\vdots 	&\vdots 	&\vdots	&\vdots 	&\vdots	&\vdots	&\vdots	&\vdots	&\vdots 	&\vdots 	&\vdots 	&\vdots &\vdots 	&\vdots 	&\vdots	&\vdots				 	\\
\D	& \D	& 0 	&\dots	& 0	& \Dp 	& 0 	& \dots 	& 0 	& \D 	& 0 	& \dots 	& 0 	& \D 	& 0 	& \dots 	& 0 	& \D 	& \dots 	& \Dp 	& \dots 	& \dots 					\\
\vdots  	&\vdots 	&\vdots 	&\vdots	&\vdots 	&\vdots	&\vdots 	&\vdots 	&\vdots	&\vdots 	&\vdots	&\vdots	&\vdots	&\vdots	&\vdots 	&\vdots 	&\vdots 	&\vdots &\vdots 	&\vdots 	&\vdots	&\vdots				 	\\
\D	& \dots 	& \D	& 0 	&\dots	& 0	& \Dp 	& 0 	& \dots 	& 0 	& \D 	& 0 	& \dots 	& 0 	& \D 	& 0 	& \dots 	& 0 	& \D 	& \dots 	& \Dp 	& \dots 	& \rightarrow &	\lambda\mu+\mu	\\ 
	&	&	&\downarrow&	&	&\downarrow&																						\\
	&	&	& \mu	&	&	& \lambda \mu+\mu																					\\								
 \end{smallmatrix}
 \right).
$$
In $T_1$ we note that for any row $h$ and any column $\mu\leq j \leq (s-1)\mu$ we have:  
\begin{equation}\label{eq:T22}
 T_1[h,j]=\D\implies T_1[h,j+\mu]=\D
\end{equation}
We recall that $T_1$ has full rank as proved in \cite{CGC-cd-art-maxbetti}.
\begin{lemma}\label{lem:subTT2}
The singleton procedure is successful for $T_1$ and thus $\mathrm{prk}(T_1)=\lambda\mu+\mu$.
\begin{proof}
See \cite{CGC-cd-art-maxbetti}, Proof of Theorem~3.1, pag. 3703.
\end{proof}
\end{lemma}
Then any matrix containing $T_1$ has rank at least $\lambda\mu +\mu$, and we obtain \eqref{eq2UUU}. If $(\mu,n)\leq \mu-1$ (which it holds if and only if $\mu \nmid n$, since $\mu\leq n$), then we consider another matrix, $T_2$, in which we collect $s$ rows of $M(\vv)$: the $\left((n-i''+k\mu)_{n}+1\right)-$th rows with $k=1,\dots,s$, which are the rows with the secondary pivot in position $k\mu$.
$$
T_2=\left(
\begin{smallmatrix}
\dots 	& 0 	& \D 	& \dots 	& \Dp 		& \dots 	& \dots 	& \dots	&\dots 	&\dots 	&\dots	&\dots	&\dots	&\dots	&\dots	&\dots	&\dots	&\dots	&\dots	&\dots	&\dots	&\dots	&\dots 	&\dots	& \dots	&\dots	&\dots	&0	& \rightarrow 	&	1	\\
\dots 	& 0 	& \D 	& 0 	& \dots 		& 0 	& \D 	& \dots 	& \Dp 	& \dots	&\dots 	&\dots 	&\dots	&\dots	&\dots	&\dots	&\dots	&\dots	&\dots	&\dots	&\dots	&\dots	&\dots 	&\dots	& \dots	&\dots	&\dots	&0	& \rightarrow 	&	2	\\
\dots 	& 0 	& \D 	& 0	& \dots 		& 0 	& \D	& 0	& \dots	& 0 	& \D 	& \dots 	& \Dp 	& \dots	&\dots 	&\dots  &\dots	&\dots	&\dots	&\dots	&\dots	&\dots	&\dots 	&\dots	& \dots	&\dots	&\dots	&0	& \rightarrow 	&	3	\\
\vdots  	&\vdots 	&\vdots 	&\vdots	&\vdots 		&\vdots	&\vdots 	&\vdots 	&\vdots	&\vdots 	&\vdots	&\vdots	&\vdots	&\vdots	&\vdots 	&\vdots 	&\vdots 	&\vdots &\vdots 	&\vdots 	&\vdots	&\vdots	&	&	&	&	&	& 	&		&		\\
\dots 	& 0 	& \D 	& 0 	& \dots 		& 0 	& \D 	& 0 	& \dots	& 0 	& \D 	& 0 	& \dots 	& 0 	& \D 	& \dots	& 0	&\dots	&0	&\D	&\dots 	&\Dp	&\dots	&\dots	& \dots	&\dots	&\dots	&0	&\rightarrow	&	s-1	\\ 
\dots 	& 0 	& \D 	& 0 	& \dots 		& 0 	& \D 	& 0 	& \dots	& 0 	& \D 	& 0 	& \dots 	& 0 	& \D 	& \dots	& 0	&\dots	&0	&\D	&0 	&\dots	&0	&\D	&\dots 	&\Dp	&\dots 	&0	&\rightarrow	&	s	\\ 
	&	&	&	&\downarrow	&	&	&	&\downarrow&	&	&	&\downarrow&	&	&	&	&	&	&	&	&\downarrow&	&	&	&\downarrow&	&	&		&		\\
	&	&	& 	&\mu		&	& 	&	&2\mu	&	&	&	&3\mu	&	&	&	&	&	& 	&	&	&(s-1)\mu&	&	&	&\mu	&	&	&		&		\\								
 \end{smallmatrix}
\right)
$$
Note that there may be some rows in common between $T_1$ and $T_2$.
\begin{lemma}
 The singleton procedure is successful for $T_2$ and thus $\mathrm{prk}(T_2)=s$. Moreover, $T_2[h]$ is the row corresponding to the singleton $T_2(h\mu)$ for $ 1\leq h \leq s$.
\begin{proof} The rows in $T_2$ correspond to the rows of matrix $T_3$ in the proof of Proposition~\ref{boundD1U}, but a shift and a permutation, so it is enough to apply Lemma~\ref{lem:subT3} and Lemma~\ref{lemma:LinAlgU}. 
\end{proof}
\end{lemma}
Our aim is to put together the rows of $T_1$ and $T_2$, obtaining a matrix $T$, and identifying a submatrix $\widetilde{T}$ of $T$, where we apply the singleton procedure. 
\[
 T=
 \left(\begin{smallmatrix}
\Dp	& 0  	&\dots 	& 0 	& \D 	& 0 	& \dots 	& 0 	& \D 	& \dots 	& \dots 	& \dots 	& \dots 	& 0 	& \dots 	& 0 	& \D 	 &\dots	& \dots 	& \dots	& \dots 	& \dots & \rightarrow	& 1	\\ 
  0	& \Dp 	& 0 	& \dots 	& 0  	& \D 	& 0 	& \dots 	& 0 	& \D 	& \dots 	& \dots 	& \dots	& \dots & 0 	& 0 	& 0 	& \D 	& \dots 	& \dots 	& \dots 	& \dots  			\\
  0	& 0	& \Dp 	& 0 	& \dots 	& 0  	& \D 	& 0 	& \dots 	& 0 	& \D 	& \dots 	& \dots 	& \dots 	& \dots 	& 0 	& 0 	& 0 	& \D 	& \dots 	& \dots 	& \dots 				\\ 
  0	& 0	& 0	& \Dp 	& 0 	& \dots 	& 0  	& \D 	& 0 	& \dots 	& 0	& \D 	& \dots 	& \dots 	& \dots 	& \dots 	& 0 	& 0 	& 0 	& \D 	& \dots 	& \dots 				\\
  0	& \dots	& \dots	& 0	& \Dp 	& 0 	& \dots 	& 0  	& \D 	& 0 	& \dots 	& 0	& \D 	& \dots 	& \dots 	& \dots 	& \dots 	& 0 	& 0 	& 0 	& \D 	& \dots 				\\
  \vdots	&\vdots	& \vdots	& \vdots	& \vdots	& \vdots	& \vdots	& \vdots	& \vdots	& \vdots	& \vdots	& \vdots	& \vdots	& \vdots	& \vdots	& \vdots	& \vdots	& \vdots	& \vdots	& \vdots	& \vdots	& \vdots				\\
  \D	& 0	& \dots	& \dots 	& \dots	& \dots	& 0	& \D	& \dots	& \Dp 	& 0 	& \dots 	& 0  	& \D 	& 0 	& \dots 	& 0	& \D 	& \dots 	& \dots 	& \dots 	& \dots 				\\
  \D 	& \D	& 0	& \dots	& \dots 	& \dots	& \dots	& 0	& \D	& \dots	& \Dp 	& 0 	& \dots 	& 0  	& \D 	& 0 	& \dots 	& 0	& \D 	& \dots 	& \dots 	& \dots 				\\  
  \D	& \dots 	& \D	& 0	& \dots	& \dots 	& \dots	& \dots	& 0	& \D	& \dots	& \Dp 	& 0 	& \dots 	& 0  	& \D 	& 0 	& \dots 	& 0	& \D 	& \dots 	& \dots 	& \rightarrow	& \lambda\mu+\mu	\\\\
 \\
 \hline
 \\
  0 	& \D 	& \dots 	& \Dp 		& \dots 	& \dots 	& \dots 	& \dots 		& \dots 	& \dots 	& \dots 	& \dots 		&\dots	& \dots 	& \dots 	& \dots 	& \dots 	& \dots 	& \dots 	& \dots	& \dots 	& \dots 	& \rightarrow	& 1	\\
  0 	& \D 	& 0 	& \dots 		& 0 	& \D 	& \dots 	& \Dp 		& \dots 	& \dots 	& \dots 	& \dots 		& \dots 	& \dots 	& \dots 	& \dots 	&\dots	& \dots 	& \dots 	& \dots 	& \dots 	& \dots 				\\
  \vdots	&\vdots	& \vdots	& \vdots		& \vdots	& \vdots	& \vdots	& \vdots		& \vdots	& \vdots	& \vdots	& \vdots		& \vdots	& \vdots	& \vdots	& \vdots	& \vdots	& \vdots	& \vdots	& \vdots	& \vdots	& \vdots				\\
  \dots 	& \dots 	& 0 	& \dots 		& 0 	& \D 	& 0 	& \dots 		& 0 	& \D 	& \dots 	& \Dp 		& \dots 	& \dots 	& \dots 	& \dots 	& \dots 	& \dots 	& \dots 	& \dots 	&\dots	& \dots 				\\
  0 	& \D 	& 0 	& \dots 		& 0 	& \D 	& \dots 	& \dots 		& \dots 	& \dots 	& 0 	& \dots 		& 0 	& \D 	& \dots 	& \Dp 	& \dots 	& \dots 	& \dots 	& \dots 	& \dots 	& \dots 	& \rightarrow	& s	\\
	&	&	&\downarrow	&	&	&	&\downarrow	&	&	&	&\downarrow	&	&	&	&\downarrow&									\\
	&	&	&\mu		&	&	&	& \lambda\mu		&	&	&	&\dots		&	&	&	& s\mu											
 \end{smallmatrix}\right)
\]
In order to do that, we use the singletons of the matrix $T_2$, removing, if necessary, some rows of $T_1$. Let $k=1, \dots, s$ and $B_{k\mu}$ be the set of the rows of $T_1$ to discard so that $T(k\mu)$ become a singleton. In other words, $B_{k\mu}=\Set{h \mid T_1[h,k\mu]=\D}$. To determine the maximal number of the discarded rows of $T_1$, we have to estimate the size of $\mathbf{B}=\cup_{k=1}^{s}B_{k\mu}$. Thanks to \eqref{eq:T22}, if $k'\leq k$ then $B_{k'\mu}\subseteq B_{k\mu}$, so $\mathbf{B}=B_{s\mu}$ and it is enough to estimate 
\begin{align*}
  \eta		&	= | \Set{h \mid T_1[h,s\mu]=\D, 1\leq h \leq \lambda\mu+\mu} |		\\
		&	= | \Set{h \mid T_1[1,s\mu-h]=\D, 0\leq h \leq \lambda\mu+\mu-1} |. 		\\
		&	= | \Set{h \mid \vv[s\mu-h]=\D, 0\leq h \leq \lambda\mu+\mu-1}|. 
  \end{align*}
Since $s\geq \lambda+1$, starting from $\vv[s\mu]$ and moving to the left of $(\lambda\mu +\mu)$ positions, we meet exactly $\lambda+1$ blocks $(0^{\mu-1}\D)$, each
contributing to $\eta$ by at most $1$, so $\eta\leq\lambda+1$.
\begin{remark}
 Note that for the computation of $\eta$ we did not need to use Lemma~\ref{lem:eta}, since this time we know exactly where the secondary pivot is, thus the determination of $\eta$ is easier.
\end{remark}
In conclusion, we have just proved that discarding at most $\lambda+1$ rows of $T$, we obtain a submatrix $\widetilde{T}$ of $T$ for which the singleton procedure is successful and we conclude: 
\[
\mathrm{prk}(T)\geq\mathrm{prk}(\widetilde{T})=\lambda\mu+\mu+s-\lambda-1. \qed  
\]
\end{proof}
\begin{example} Let $C$ be a cyclic code of length $n=27$, with defining set $S_C$ satisfying the assumptions of Proposition~\ref{boundD2U} with parameters $\mu=4$, $\lambda=2$, $s=5$. We want to prove that by Proposition~\ref{boundD2U} the distance of the code $C$ is at least $d\geq 4\cdot 2+4+4-2-1=13.$ 
 Let $\vv\in\mathcal{A}(R(n,S_C))$, then we can suppose $\vv[1]=\Dp$ and  $i''=18$ or $i''=19$, otherwise the bound is trivially satisfied.\\
 \emph{ Case 1: $i''=18$}, $\vv=\Dp000\D000\D000\D000\D\Dp\D00000000$.
 \[
  \begin{smallmatrix}
 \Dp	& 0  	& 0 	& 0 	& \D 	& 0 	& 0 	& 0 	& \D 	& 0 	& 0 	& 0 	& \D 	& 0 	& 0 	& 0 	& \D 	 & \Dp 	& \D 	& 0 	& 0 	& 0 	& 0 	& 0 	& 0 	& 0 	& 0 	&\rightarrow	&	\mbox{\scriptsize{$5$-th s-deletion}}  	\\ 
  0	& \Dp 	& 0 	& 0 	& 0  	& \D 	& 0 	& 0 	& 0 	& \D 	& 0 	& 0 	& 0	& \D  	& 0 	& 0 	& 0 	& \D 	& \Dp 	& \D 	& 0 	& 0 	& 0 	& 0 	& 0 	& 0 	& 0  	&\rightarrow	&	\mbox{\scriptsize{$6$-th s-deletion}}	\\
  0	& 0	& \Dp 	& 0 	& 0 	& 0  	& \D 	& 0 	& 0 	& 0 	& \D 	& 0 	& 0 	& 0 	& \D 	& 0 	& 0 	& 0 	& \D 	& \Dp 	& \D 	& 0 	& 0 	& 0 	& 0 	& 0 	& 0 	&\rightarrow	&	\mbox{\scriptsize{$7$-th s-deletion}}	\\
  \rlap{\rule[0.8mm]{0.90\textwidth}{0.4mm}}
  0	& 0	& 0	& \Dp 	& 0 	& 0 	& 0  	& \D 	& 0 	& 0 	& 0	& \D 	& 0 	& 0 	& 0 	& \D 	& 0 	& 0 	& 0 	& \D 	& \Dp 	& \D 	& 0 	& 0 	& 0 	& 0 	& 0	&\rightarrow	&	\mathbf{DISCARDED}  			\\
  0	& 0	& 0	& 0	& \Dp 	& 0 	& 0 	& 0  	& \D 	& 0 	& 0 	& 0	& \D 	& 0 	& 0 	& 0 	& \D 	& 0 	& 0 	& 0 	& \D 	& \Dp 	& \D 	& 0 	& 0 	& 0 	& 0 	&\rightarrow	&	\mbox{\scriptsize{$8$-th s-deletion}}	\\
  0	& 0	& 0	& 0	& 0	& \Dp 	& 0 	& 0 	& 0  	& \D 	& 0 	& 0 	& 0	& \D 	& 0 	& 0 	& 0 	& \D 	& 0 	& 0 	& 0 	& \D 	& \Dp 	& \D 	& 0 	& 0 	& 0 	&\rightarrow	&	\mbox{\scriptsize{$9$-th s-deletion}}	\\
  0 	& 0	& 0	& 0	& 0	& 0	& \Dp 	& 0 	& 0 	& 0  	& \D 	& 0 	& 0 	& 0	& \D 	& 0 	& 0 	& 0 	& \D 	& 0 	& 0 	& 0 	& \D 	& \Dp 	& \D 	& 0 	& 0 	&\rightarrow	&	\mbox{\scriptsize{$10$-th s-deletion}}  	\\
  \rlap{\rule[0.8mm]{0.90\textwidth}{0.4mm}}
  0	& 0 	& 0	& 0	& 0	& 0	& 0	& \Dp 	& 0 	& 0 	& 0  	& \D 	& 0 	& 0 	& 0	& \D 	& 0 	& 0 	& 0 	& \D 	& 0 	& 0 	& 0 	& \D 	& \Dp 	& \D 	& 0 	&\rightarrow	&	\mathbf{DISCARDED}				\\
  0	& 0	& 0 	& 0	& 0	& 0	& 0	& 0	& \Dp 	& 0 	& 0 	& 0  	& \D 	& 0 	& 0 	& 0	& \D 	& 0 	& 0 	& 0 	& \D 	& 0 	& 0 	& 0 	& \D 	& \Dp 	& \D 	&\rightarrow	&	\mbox{\scriptsize{$11$-th s-deletion}}	\\
  \D	& 0	& 0	& 0 	& 0	& 0	& 0	& 0	& 0	& \Dp 	& 0 	& 0 	& 0  	& \D 	& 0 	& 0 	& 0	& \D 	& 0 	& 0 	& 0 	& \D 	& 0 	& 0 	& 0 	& \D 	& \D 	&\rightarrow	&	\mbox{\scriptsize{$12$-th s-deletion}}	\\
  \Dp 	& \D	& 0	& 0	& 0 	& 0	& 0	& 0	& 0	& 0	& \Dp 	& 0 	& 0 	& 0  	& \D 	& 0 	& 0 	& 0	& \D 	& 0 	& 0 	& 0 	& \D 	& 0 	& 0 	& 0 	& \D 	&\rightarrow	&	\mbox{\scriptsize{$13$-th s-deletion}}  	\\  
  \rlap{\rule[0.8mm]{0.90\textwidth}{0.4mm}}
  \D	& \Dp 	& \D	& 0	& 0	& 0 	& 0	& 0	& 0	& 0	& 0	& \Dp 	& 0 	& 0 	& 0  	& \D 	& 0 	& 0 	& 0	& \D 	& 0 	& 0 	& 0 	& \D 	& 0 	& 0 	& 0 	&\rightarrow	&	\mathbf{DISCARDED}				\\
  \\  
  0 	& 0 	& \D 	& \Dp 	& \D 	& 0 	& 0 	& 0 	& 0 	& 0 	& 0 	& 0 	& 0 	&\Dp	& 0 	& 0 	& 0 	& \D 	& 0 	& 0 	& 0 	& \D 	& 0 	& 0 	& 0 	& \D 	& 0	&\rightarrow	&	\mbox{\scriptsize{$1$-st s-deletion}} 	\\
  0	& 0 	& \D 	& 0 	& 0 	& 0 	& \D 	& \Dp 	& \D 	& 0 	& 0 	& 0 	& 0 	& 0 	& 0 	& 0 	& 0 	&\Dp	& 0 	& 0 	& 0 	& \D 	& 0 	& 0 	& 0 	& \D 	& 0 	&\rightarrow	&	\mbox{\scriptsize{$2$-nd s-deletion}}	\\
  0	& 0 	& \D 	& 0 	& 0 	& 0 	& \D 	& 0 	& 0 	& 0 	& \D 	& \Dp 	& \D 	& 0 	& 0 	& 0 	& 0 	& 0 	& 0 	& 0 	& 0 	&\Dp	& 0 	& 0 	& 0 	& \D 	& 0 	&\rightarrow	&	\mbox{\scriptsize{$3$-rd s-deletion}}	\\
  0	& 0 	& \D 	& 0 	& 0 	& 0 	& \D 	& 0 	& 0 	& 0 	& \D 	& 0 	& 0 	& 0 	& \D 	& \Dp 	& \D 	& 0 	& 0 	& 0 	& 0 	& 0 	& 0 	& 0 	& 0 	&\Dp	& 0 	&\rightarrow	&	\mbox{\scriptsize{$4$-th s-deletion}}   	\\
  \end{smallmatrix}
 \]
 \emph{ Case 2: $i''=19$}, $\vv=\Dp000\D000\D000\D000\D\D\Dp00000000$.
 
 \[
  \begin{smallmatrix}
 \Dp	& 0  	& 0 	& 0 	& \D 	& 0 	& 0 	& 0 	& \D 	& 0 	& 0 	& 0 	& \D 	& 0 	& 0 	& 0 	& \D 	 & \D 	& \Dp 	& 0 	& 0 	& 0 	& 0 	& 0 	& 0 	& 0 	& 0 	&\rightarrow	&	\mbox{\scriptsize{$5$-th s-deletion}}  	\\ 
  0	& \Dp 	& 0 	& 0 	& 0  	& \D 	& 0 	& 0 	& 0 	& \D 	& 0 	& 0 	& 0	& \D  	& 0 	& 0 	& 0 	& \D 	& \D 	& \Dp 	& 0 	& 0 	& 0 	& 0 	& 0 	& 0 	& 0  	&\rightarrow	&	\mbox{\scriptsize{$6$-th s-deletion}}	\\
  0	& 0	& \Dp 	& 0 	& 0 	& 0  	& \D 	& 0 	& 0 	& 0 	& \D 	& 0 	& 0 	& 0 	& \D 	& 0 	& 0 	& 0 	& \D 	& \D 	& \Dp 	& 0 	& 0 	& 0 	& 0 	& 0 	& 0 	&\rightarrow	&	\mbox{\scriptsize{$7$-th s-deletion}}	\\
  \rlap{\rule[0.8mm]{0.90\textwidth}{0.4mm}}
  0	& 0	& 0	& \Dp 	& 0 	& 0 	& 0  	& \D 	& 0 	& 0 	& 0	& \D 	& 0 	& 0 	& 0 	& \D 	& 0 	& 0 	& 0 	& \D 	& \D 	& \Dp 	& 0 	& 0 	& 0 	& 0 	& 0	&\rightarrow	&	\mathbf{DISCARDED}  			\\
  0	& 0	& 0	& 0	& \Dp 	& 0 	& 0 	& 0  	& \D 	& 0 	& 0 	& 0	& \D 	& 0 	& 0 	& 0 	& \D 	& 0 	& 0 	& 0 	& \D 	& \D 	& \Dp 	& 0 	& 0 	& 0 	& 0 	&\rightarrow	&	\mbox{\scriptsize{$8$-th s-deletion}}	\\
  0	& 0	& 0	& 0	& 0	& \Dp 	& 0 	& 0 	& 0  	& \D 	& 0 	& 0 	& 0	& \D 	& 0 	& 0 	& 0 	& \D 	& 0 	& 0 	& 0 	& \D 	& \D 	& \Dp 	& 0 	& 0 	& 0 	&\rightarrow	&	\mbox{\scriptsize{$9$-th s-deletion}}	\\
  0 	& 0	& 0	& 0	& 0	& 0	& \Dp 	& 0 	& 0 	& 0  	& \D 	& 0 	& 0 	& 0	& \D 	& 0 	& 0 	& 0 	& \D 	& 0 	& 0 	& 0 	& \D 	& \D 	& \Dp 	& 0 	& 0 	&\rightarrow	&	\mbox{\scriptsize{$10$-th s-deletion}}  	\\
  \rlap{\rule[0.8mm]{0.90\textwidth}{0.4mm}}
  0	& 0 	& 0	& 0	& 0	& 0	& 0	& \Dp 	& 0 	& 0 	& 0  	& \D 	& 0 	& 0 	& 0	& \D 	& 0 	& 0 	& 0 	& \D 	& 0 	& 0 	& 0 	& \D 	& \D 	& \Dp 	& 0 	&\rightarrow	&	\mathbf{DISCARDED}				\\
  0	& 0	& 0 	& 0	& 0	& 0	& 0	& 0	& \Dp 	& 0 	& 0 	& 0  	& \D 	& 0 	& 0 	& 0	& \D 	& 0 	& 0 	& 0 	& \D 	& 0 	& 0 	& 0 	& \D 	& \D 	& \Dp 	&\rightarrow	&	\mbox{\scriptsize{$11$-th s-deletion}}	\\
  \Dp	& 0	& 0	& 0 	& 0	& 0	& 0	& 0	& 0	& \Dp 	& 0 	& 0 	& 0  	& \D 	& 0 	& 0 	& 0	& \D 	& 0 	& 0 	& 0 	& \D 	& 0 	& 0 	& 0 	& \D 	& \D 	&\rightarrow	&	\mbox{\scriptsize{$12$-th s-deletion}}	\\
  \D 	& \Dp	& 0	& 0	& 0 	& 0	& 0	& 0	& 0	& 0	& \Dp 	& 0 	& 0 	& 0  	& \D 	& 0 	& 0 	& 0	& \D 	& 0 	& 0 	& 0 	& \D 	& 0 	& 0 	& 0 	& \D 	&\rightarrow	&	\mbox{\scriptsize{$13$-th s-deletion}}  	\\  
  \rlap{\rule[0.8mm]{0.90\textwidth}{0.4mm}}
  \D	& \D 	& \Dp	& 0	& 0	& 0 	& 0	& 0	& 0	& 0	& 0	& \Dp 	& 0 	& 0 	& 0  	& \D 	& 0 	& 0 	& 0	& \D 	& 0 	& 0 	& 0 	& \D 	& 0 	& 0 	& 0 	&\rightarrow	&	\mathbf{DISCARDED}				\\
  \\  
  0 	& \D 	& \D 	& \Dp 	& 0 	& 0 	& 0 	& 0 	& 0 	& 0 	& 0 	& 0 	&\Dp	& 0 	& 0 	& 0 	& \D 	& 0 	& 0 	& 0 	& \D 	& 0 	& 0 	& 0 	& \D 	& 0	& 0		&\rightarrow	&	\mbox{\scriptsize{$1$-st s-deletion}} 	\\
  0 	& \D 	& 0 	& 0 	& 0 	& \D 	& \D 	& \Dp 	& 0 	& 0 	& 0 	& 0 	& 0 	& 0 	& 0 	& 0 	&\Dp	& 0 	& 0 	& 0 	& \D 	& 0 	& 0 	& 0 	& \D 	& 0 	& 0		&\rightarrow	&	\mbox{\scriptsize{$2$-nd s-deletion}}	\\
  0 	& \D 	& 0 	& 0 	& 0 	& \D 	& 0 	& 0 	& 0 	& \D 	& \D 	& \Dp 	& 0 	& 0 	& 0 	& 0 	& 0 	& 0 	& 0 	& 0 	&\Dp	& 0 	& 0 	& 0 	& \D 	& 0 	& 0		&\rightarrow	&	\mbox{\scriptsize{$3$-rd s-deletion}}	\\
  0 	& \D 	& 0 	& 0 	& 0 	& \D 	& 0 	& 0 	& 0 	& \D 	& 0 	& 0 	& 0 	& \D 	& \D 	& \Dp 	& 0 	& 0 	& 0 	& 0 	& 0 	& 0 	& 0 	& 0 	&\Dp	& 0 	& 0		&\rightarrow	&	\mbox{\scriptsize{$4$-th s-deletion}}   	\\
  \end{smallmatrix}
 \]

\end{example}

We summarize the results of Proposition~\ref{boundD1U} and Proposition~\ref{boundD2U} in one statement, called bound C.
%

\begin{theorem}[bound $\texttt{C}$]\label{boundD}
Let $C$ be a $[n,k,d]$ cyclic code with defining set $S_C$. Suppose that there are $\ell, \ m, \ r, \ s, \ \rho \in\mathbb{N} $, $1\leq m\leq \ell$, $s\geq 1$, $\rho\geq 1$ such that 
\begin{align*}
((0)^\ell (\D)^r)((0)^{m}(\D)^r)^{s} \preccurlyeq R(n,S_C)^\rho		&&	\mbox{or}	&&	((\D)^r (0)^{m})^{s}((\D)^r (0)^\ell) \preccurlyeq R(n,S_C)^\rho.	\\ 
\end{align*}
Then:
\begin{itemize}
 \item if $(m+r,n)\leq m$:
\[
d\geq \ell +1 +s-r \left\lfloor\frac{\ell}{m+r}\right\rfloor-\max\Set{(\ell)_{m+r}-\lambda,0}; 
\]
 \item otherwise:
\[
d\geq \ell +1 .
\]
\end{itemize}

In the particular case that, $\ell=\lambda\mu$, $m=\mu-1$, $s\geq \lambda+1$ and $r=1$ for some $\mu$ and $\lambda$, we also have: 
\begin{itemize}
 \item  $ d\geq \mu \lambda+\mu+s-\lambda-1$, if $\mu\nmid n$
 \item  $ d\geq \mu \lambda+\mu$, otherwise.
\end{itemize}
\end{theorem}


\section{Computational costs and results}
\label{results}

As explained in Remark~\ref{rem:D1} and in Remark~\ref{rem:D2} bound $\texttt{C}$ is both a generalization of the HT bound and the BS bound (except when $\mu | n$) and so it is sharper and tighter. The relation between our bound and the Roos bound is not clear: sometimes our bound is sharper and tighter than Roos's but for other codes it is the opposite. However, from the computed codes it appears that bound $\texttt{C}$ is tighter than the Roos bound overall.
Although the BS bound sometimes beats the Roos bound, in the majority of computed cases the Roos bound is better, as reported in \cite{CGC-cd-art-maxbetti} and checked by us. Bound $\texttt{C}$ is the first polynomial-time bound outperforming the Roos bound on a significant sample of codes.\\ 
As regards computational costs, bound $\texttt{C}$ requires at most:
\begin{itemize}
  \item $n$ operations for $i_0$
  \item $n$ operations for $\ell$,  
  \item $n$ operations for $ m $,
  \item $n$ operations for $ r $,
  \item $n$ operations for $ s $
\end{itemize}  
and so it costs $O(n^5)$ which is slightly more than the Roos bound which needs $O(n^4)$, in fact the latter requires at most:
\begin{itemize}
  \item $n$ operations for $i_0$,  
  \item $n$ operations for $ m $,
  \item $n$ operations for $ r $,
  \item $n$ operations for $ s $
\end{itemize}  
while the other bounds cost less: $\BCH$-$O(n^2)$, HT-$O(n^3)$, bound BS-$O(n^{2.5})$. 
We tested all cyclic codes in the following range:  on $\mathbb{F}_{2}$ with $15\leq n\leq 125$, on $\mathbb{F}_{3}$ with $8\leq n\leq 79$ and $82\leq n\leq 89$, on $\mathbb{F}_{5}$ with $8\leq n\leq 61$, on
$\mathbb{F}_{7}$ with $8\leq n\leq 47$. 
We have chosen the largest ranges that we could compute in a reasonable time. In Appendix, Table~\ref{tab1}-~\ref{tab2}-~\ref{tab3}-~\ref{tab4}-~\ref{tab5}-~\ref{tab6} give in detail the results obtained for each characteristic. We write $\BCH$ for the BCH bound, $\HT$ for the HT bound, $\mathrm{BS}$ for the BS bound,$\mathrm{RS}$ for the Roos bound and $\mathrm{BC}$ for the bound $\texttt{C}$. 

Since all the bounds that we consider are sharper than the $\BCH$ bound, clearly they are tight for all cyclic codes in which the $\BCH$ bound is already tight. Thus, it is interesting to consider the only cases when the $\HT$, BS, Roos and $\texttt{C}$ bounds are tight and the $\BCH$ bound is not.       
The following table is composed of two different parts. 
In the first part we report: in the first row the number of checked codes, in the second row the number of these for which the $\BCH$ bound is tight. 
In the second part of the table, each row corresponds to a specific bound. For each row we report the number of codes for which the bound is tight and the $\BCH$ bound is not.
\begin{table}[h]
\caption{Bound tightness}
\label{tab:riassuntiva}
$$
\begin{array}{ccccc|c}
\hline
			&\mathbb{F}_{2}		&\mathbb{F}_{3}		&\mathbb{F}_{5}	&\mathbb{F}_{7}	& \text{total}			\\
\hline
\text{number of codes}		&70488		&93960			&1163176		&106804		&1434428		\\

\text{BCH}			&59296			&77584				&1011957		&93108		&1241945		\\
\hline
\text{HT}			&661			&1042				&12058			&2603			&16364			\\

\text{BS}			&233			&831				&11436			&2413			&14913			\\

\text{ROOS}			&{\bf 1178}		&1793				&17673			&2987			&23631			\\

\text{bound $\texttt{C}$}	 	&886			&{\bf 1811}			&{ \bf 20147}		&{\bf 4155}		&{\bf 26999}			\\
\hline
\end{array}
$$
\end{table}


\section*{Acknowledgements}
These bounds appear in the 2010 Master's thesis of the first author \cite{CGC-alg-tesi2-piva10}, who thanks his supervisor (the second author), and were partially presented at the conference ``Trends in Coding Theory '' in Ascona (October-November~2012). 
\clearpage


\section*{Appendix}
\begin{figure}[h!]
\begin{minipage}[h!]{0.47\textwidth}
\centering
\begin{tiny}
 $
 \begin{array}{|c|c|c|c|c|c|c|c|}
 \hline 
    n    & N_{codes}  	&  \text{BCH} 	&   \text{HT} 	&  \text{BS} 	&     \text{RS} 	&   \text{BC} 	\\
\hline
 15 &       32 &       30 &       32 &       30 &       32 &               32 \\ 
\hline
   17 &        8 &        5 &        8 &        5 &        8 &               8 \\ 
\hline
   19 &        4 &        4 &        4 &        4 &        4 &                4 \\ 
\hline
   21 &       64 &       52 &       54 &       52 &       58 &              54 \\ 
\hline
   23 &        8 &        4 &        4 &        4 &        4 &                4 \\ 
\hline
   25 &        8 &        8 &        8 &        8 &        8 &               8 \\ 
\hline
   27 &       16 &       16 &       16 &       16 &       16 &               16 \\ 
\hline
   29 &        4 &        4 &        4 &        4 &        4 &               4 \\ 
\hline
   31 &      128 &       46 &       96 &       46 &       96 &              96 \\ 
\hline
   33 &       32 &       21 &       26 &       21 &       26 &              26 \\ 
\hline
   35 &       64 &       40 &       42 &       40 &       48 &              44 \\ 
\hline
   37 &        4 &        4 &        4 &        4 &        4 &               4 \\ 
\hline
   39 &       32 &       18 &       20 &       18 &       20 &              20 \\ 
\hline
   41 &        8 &        4 &        4 &        4 &        4 &               4 \\ 
\hline
   43 &       16 &        6 &       10 &        6 &       11 &              10 \\ 
\hline
   45 &      256 &      187 &      222 &      189 &      228 &           224 \\ 
\hline
   47 &        8 &        4 &        4 &        4 &        4 &               4 \\ 
\hline
   49 &       32 &       32 &       32 &       32 &       32 &             32 \\ 
\hline
   51 &      256 &       90 &      146 &       98 &      146 &            150 \\ 
\hline
   53 &        4 &        4 &        4 &        4 &        4 &               4 \\ 
\hline
   55 &       32 &       16 &       20 &       16 &       20 &              20 \\ 
\hline
   57 &       32 &       20 &       24 &       20 &       24 &              24 \\ 
\hline
   59 &        4 &        4 &        4 &        4 &        4 &               4 \\ 
\hline
   61 &        4 &        4 &        4 &        4 &        4 &              4 \\ 
\hline
   63 &     8192 &     2238 &     4210 &     2401 &     4346 &           4280 \\ 
\hline
   65 &      128 &       36 &       74 &       36 &       78 &             74 \\ 
\hline
   67 &        4 &        4 &        4 &        4 &        4 &              4 \\ 
\hline
   69 &       64 &       22 &       24 &       22 &       24 &             24 \\ 
 \hline 
\end{array}
$
\end{tiny}
\captionof{table}{\label{tab1}Tightness $\FF_2$, $15\leq n\leq 69$}
\end{minipage}
\hfill
\begin{minipage}[h!]{0.47\textwidth}
\centering
\begin{tiny}
$
\begin{array}{|c|c|c|c|c|c|c|c|}
 \hline 
    n    & N_{codes}  	&  \text{BCH} 	&   \text{HT} 	&  \text{BS} 	&     \text{RS} 	&   \text{BC} 	\\
\hline
   71 &        8 &        4 &        4 &        4 &        4 &              4 \\ 
\hline
   73 &      512 &       37 &      104 &       39 &      117 &            106 \\ 
\hline
   75 &      256 &      220 &      252 &      220 &      254 &            252 \\ 
\hline
   77 &       64 &       42 &       44 &       42 &       44 &             44 \\ 
\hline
   79 &        8 &        4 &        4 &        4 &        4 &              4 \\ 
\hline
   81 &       32 &       32 &       32 &       32 &       32 &             32 \\ 
\hline
   83 &        4 &        4 &        4 &        4 &        4 &              4 \\ 
\hline
   85 &     4096 &      547 &     1124 &      571 &     1141 &           1132 \\ 
\hline
   87 &       32 &       18 &       20 &       18 &       20 &             20 \\ 
\hline
   89 &      512 &       20 &       56 &       20 &       56 &             56 \\ 
\hline
   91 &     1024 &      277 &      435 &      277 &      436 &            435 \\ 
\hline
   93 &    16384 &     1388 &     3268 &     1424 &     3360 &           3286 \\ 
\hline
   95 &       32 &       18 &       20 &       18 &       20 &             20 \\ 
\hline
   97 &        8 &        4 &        4 &        4 &        4 &              4 \\ 
\hline
   99 &      256 &      105 &      166 &      106 &      171 &            166 \\ 
 \hline 
   101 &        4 &        4 &        4 &        4 &        4 &             4 \\
   \hline 
  103 &        8 &        4 &        4 &        4 &        4 &              4 \\ 
  \hline 
  105 &    32768 &     7939 &    11446 &     8420 &    12325 &          11796 \\ 
  \hline 
  107 &        4 &        4 &        4 &        4 &        4 &             4 \\ 
  \hline 
  109 &       16 &        4 &        4 &        4 &        4 &             4 \\ 
  \hline 
  111 &       32 &       18 &       20 &       22 &       20 &            24 \\ 
  \hline 
  113 &       32 &        4 &        4 &        4 &        4 &             4 \\ 
  \hline 
  115 &       64 &       24 &       26 &       24 &       26 &            26 \\ 
  \hline 
  117 &     4096 &      637 &     1075 &      714 &     1110 &        1099 \\ 
  \hline 
  119 &      512 &      170 &      212 &      170 &      213 &           212 \\ 
  \hline 
  121 &        8 &        8 &        8 &        8 &        8 &             8 \\ 
  \hline 
  123 &      256 &       52 &       62 &       60 &       62 &            66 \\ 
  \hline 
  125 &       16 &       16 &       16 &       16 &       16 &           16 \\
\hline
\end{array}
$
\end{tiny}

\captionof{table}{\label{tab2}Tightness $\FF_2$, $71\leq n\leq 125$}
\end{minipage}
\end{figure}

\begin{figure}[h!]
\begin{minipage}[h!]{0.47\textwidth}
\centering
\begin{tiny}
$
\begin{array}{|c|c|c|c|c|c|c|c|}
  \hline
   n    & N_{codes}  	&  \text{BCH} 	&   \text{HT} 	&  \text{BS} 	&     \text{RS} 	&   \text{BC} 	\\
\hline
     8 &       32 &       30 &       32 &       30 &       32 &          32 \\ 
\hline
   10 &       16 &       16 &       16 &       16 &       16 &           16 \\ 
\hline
   11 &        8 &        4 &        4 &        4 &        4 &            4 \\ 
\hline
   13 &       32 &       19 &       26 &       19 &       27 &            26 \\ 
\hline
   14 &       16 &       16 &       16 &       16 &       16 &            16 \\ 
\hline
   16 &      128 &      112 &      118 &      112 &      120 &           118 \\ 
\hline
   17 &        4 &        4 &        4 &        4 &        4 &             4 \\ 
\hline
   19 &        4 &        4 &        4 &        4 &        4 &             4 \\ 
\hline
   20 &      128 &       90 &      102 &      100 &      104 &           110 \\ 
\hline
   22 &       64 &       24 &       24 &       32 &       24 &            32 \\ 
\hline
   23 &        8 &        4 &        4 &        4 &        4 &             4 \\ 
\hline
   25 &        8 &        8 &        8 &        8 &        8 &             8 \\ 
\hline
   26 &     1024 &      321 &      514 &      377 &      545 &           546 \\ 
\hline
   28 &      128 &       94 &      116 &       96 &      120 &           120 \\ 
\hline
   29 &        4 &        4 &        4 &        4 &        4 &             4 \\ 
\hline
   31 &        4 &        4 &        4 &        4 &        4 &             4 \\ 
\hline
   32 &      512 &      410 &      464 &      414 &      472 &           464 \\ 
\hline
   34 &       16 &       16 &       16 &       16 &       16 &            16 \\ 
\hline
   35 &       32 &       16 &       18 &       16 &       20 &            18 \\ 
\hline
   37 &        8 &        4 &        4 &        4 &        4 &             4 \\ 
\hline
   38 &       16 &       16 &       16 &       16 &       16 &            16 \\ 
\hline
   40 &     8192 &     3170 &     4344 &     3570 &     4478 &          4614 \\ 
\hline
   41 &       64 &        9 &       29 &        9 &       30 &            29 \\ 
\hline
   43 &        4 &        4 &        4 &        4 &        4 &             4 \\ 
\hline
   44 &      512 &      208 &      216 &      236 &      218 &           244 \\ 
\hline
   46 &       64 &       24 &       24 &       24 &       24 &            24 \\ 
\hline
   47 &        8 &        4 &        4 &        4 &        4 &             4 \\ 
\hline
   49 &        8 &        8 &        8 &        8 &        8 &              8 \\ 
\hline
   50 &       64 &       64 &       64 &       64 &       64 &             64 \\ 
\hline
   52 &    32768 &     7157 &    11452 &     8281 &    12339 &          12150 \\ 
\hline
\end{array}
$
\end{tiny}
\captionof{table}{\label{tab3}Tightness $\FF_3$, $8\leq n\leq 52$}
\end{minipage}
\hfill
\begin{minipage}[h!]{0.47\textwidth}
\centering
\begin{tiny}
$
\begin{array}{|c|c|c|c|c|c|c|c|}
\hline
   n    & N_{codes}  	&  \text{BCH} 	&   \text{HT} 	&  \text{BS} 	&     \text{RS} 	&   \text{BC} 	\\
\hline
   53 &        4 &        4 &        4 &        4 &        4 &              4 \\ 
\hline
   55 &       64 &       20 &       22 &       20 &       24 &             22 \\ 
\hline
   56 &     8192 &     3168 &     4368 &     3414 &     4440 &           4466 \\ 
\hline
   58 &       16 &       16 &       16 &       16 &       16 &             16 \\ 
\hline
   59 &        8 &        4 &        4 &        4 &        4 &              4 \\ 
\hline
   61 &      128 &        5 &       10 &        5 &       11 &             10 \\ 
\hline
   62 &       16 &       16 &       16 &       16 &       16 &             16 \\ 
\hline
   64 &     2048 &     1640 &     1866 &     1652 &     1916 &           1870 \\ 
\hline
   65 &     1024 &      211 &      324 &      211 &      351 &            324 \\ 
\hline
   67 &       16 &        4 &        4 &        4 &        4 &              4 \\ 
\hline
   68 &      128 &       76 &       88 &       76 &       88 &             88 \\ 
\hline
   70 &     1024 &      422 &      454 &      450 &      464 &            490 \\ 
\hline
   71 &        8 &        4 &        4 &        4 &        4 &              4 \\ 
\hline
   73 &      128 &        5 &       10 &        5 &       10 &             10 \\ 
\hline
   74 &       64 &       28 &       32 &       28 &       32 &             32 \\ 
\hline
   76 &      128 &       92 &      112 &       92 &      112 &            112 \\ 
\hline
   77 &       64 &       20 &       22 &       20 &       22 &             22 \\ 
\hline
   79 &        4 &        4 &        4 &        4 &        4 &              4 \\ 
\hline 
    82 &     4096 &      303 &      799 &      303 &      798 &           799 \\
\hline 
   83 &        8 &        4 &        4 &        4 &        4 &              4 \\ 
\hline
   85 &      128 &       30 &       36 &       30 &       40 &             36 \\ 
\hline
   86 &       16 &       16 &       16 &       16 &       16 &             16 \\ 
\hline
   88 &    32768 &     8952 &    11484 &     9928 &    11866 &          12042 \\ 
\hline
   89 &        4 &        4 &        4 &        4 &        4 &              4 \\ 
\hline 
   92 &      512 &      196 &      204 &      196 &      204 &            204 \\
\hline
   94 &       64 &       24 &       24 &       24 &       24 &             24 \\
\hline
   95 &       32 &       18 &       20 &       18 &       20 &             20 \\
\hline
   97 &       8  &       4 &         4 &        4 &       4 &              4  \\
\hline
    98 &       64 &       64 &       64 &       64 &       64 &            64 \\
\hline
\end{array}
$
\end{tiny}
\captionof{table}{\label{tab4}Tightness $\FF_3$, $53\leq n\leq 98$, $n\neq 80$, $n\neq 91$ }
\end{minipage}
\end{figure}

\begin{figure}[h!]
\begin{minipage}[b]{0.47\textwidth}
\centering
\begin{tiny}
$
\begin{array}{|c|c|c|c|c|c|c|c|}
  \hline
   n    & N_{codes}  	&  \text{BCH} 	&   \text{HT} 	&  \text{BS} 	&     \text{RS} 	&       \text{BC} 	\\
   \hline	
    8 	&       32 	&       26 	&      32 	&       26 	&     32 	     	&   32 	\\
   \hline
    9 	&       32 	&       32 	&       32 	&       32 	&       32 	     	&       32 	\\
   \hline
   10 &       16 &       16 &       16 &       16 &       16 &               16 \\
   \hline
   11 &        4 &        4 &        4 &        4 &        4 &                4 \\
   \hline
   12 &      512 &      458 &      488 &      482 &      488 &             500   \\
   \hline
   13 &        4 &        4 &        4 &        4 &        4 &               4 \\
   \hline
   15 &       64 &       58 &       64 &       58 &       64 &              64 \\
   \hline
   16 &      512 &      218 &      326 &      250 &      336 &             342 \\
   \hline
   17 &        4 &        4 &        4 &        4 &        4 &               4 \\
   \hline
   18 &     1024 &      952 &      988 &      988 &      988 &            1012 \\
   \hline
   19 &      128 &       14 &       28 &       18 &       28 &              28 \\
   \hline
   20 &      128 &       82 &       94 &       88 &       96 &              98 \\
   \hline
   22 &       16 &       16 &       16 &       16 &       16 &              16 \\
   \hline
   23 &        4 &        4 &        4 &        4 &        4 &               4 \\
   \hline
   24 &    32768 &    15416 &    21794 &    17762 &    21836 &           22976 \\
   \hline
   25 &      128 &       28 &       72 &       29 &       74 &              72 \\
   \hline
   26 &       16 &       16 &       16 &       16 &       16 &              16 \\
   \hline
   27 &      128 &      128 &      128 &      128 &      128 &             128 \\
   \hline
   29 &       32 &        4 &        4 &        4 &        4 &               4 \\
   \hline
   30 &     4096 &     2614 &     2890 &     3046 &     2914 &            3323 \\
   \hline
   31 &        8 &        4 &        4 &        4 &        4 &               4 \\
   \hline
   32 &     8192 &     2258 &     3518 &     2480 &     3652 &            3638 \\
   \hline
   33 &       64 &       58 &       64 &       58 &       64 &              64 \\
   \hline
   34 &       16 &       16 &       16 &       16 &       16 &              16 \\
   \hline
   36 &    32768 &    25346 &    27860 &    27890 &    28124 &           29204 \\
   \hline
   37 &       32 &        4 &        4 &        4 &        4 &               4 \\
   \hline
   38 &    16384 &      762 &     1610 &      946 &     1746 &            1730 \\
   \hline
   39 &       64 &       58 &       64 &       58 &       64 &              64 \\
   \hline
   40 &     8192 &     2664 &     3598 &     2952 &     3696 &            3746 \\
   \hline
   41 &        4 &        4 &        4 &        4 &        4 &               4 \\
   \hline
   43 &      256 &        4 &        6 &        4 &        6 &               6 \\
   \hline
   44 &      128 &       84 &       96 &       84 &       98 &              96 \\
   \hline
   45 &     1024 &      763 &      850 &      763 &      856 &             850 \\
   \hline
   46 &       16 &       16 &       16 &       16 &       16 &              16 \\
   \hline
   47 &        8 &        4 &        4 &        4 &        4 &               4 \\
  \hline
\end{array}
$
\end{tiny}
\captionof{table}{\label{tab5}Tightness $\FF_7$, $8\leq n\leq 47$}
\end{minipage}
\hfill
\begin{minipage}[b]{0.47\textwidth}
\centering
\begin{tiny}
$
\begin{array}{|c|c|c|c|c|c|c|c|}
  \hline
   n    & N_{codes}  	&  \text{BCH} 	&   \text{HT} 	&  \text{BS} 	&     \text{RS} 	&      \text{BC} 	\\
\hline 
     8 &       64 &       60 &       64 &       60 &       64 &              64 \\ 
\hline
    9 &        8 &        8 &        8 &        8 &         8 &               8 \\ 
\hline
   11 &        8 &        4 &        4 &        4 &        4 &                4 \\ 
\hline
   12 &      256 &      204 &      220 &      228 &      224 &            236 \\ 
\hline
   13 &       16 &        7 &       14 &        8 &       14 &              14 \\ 
\hline
   14 &       16 &       16 &       16 &       16 &       16 &              16 \\ 
\hline
   16 &      256 &      240 &      252 &      240 &      256 &            252 \\ 
\hline
   17 &        4 &        4 &        4 &        4 &        4 &                4 \\ 
\hline
   18 &       64 &       64 &       64 &       64 &       64 &              64 \\ 
\hline
   19 &        8 &        4 &        4 &        4 &        4 &                4 \\ 
\hline
   21 &       32 &       20 &       24 &       20 &       24 &              24 \\ 
\hline
   22 &       64 &       24 &       24 &       32 &       24 &              32 \\ 
\hline
   23 &        4 &        4 &        4 &        4 &        4 &               4 \\ 
\hline
   24 &    16384 &     7264 &    10280 &     8276 &    10560 &        10720 \\ 
\hline
   26 &      256 &       81 &      156 &       92 &      156 &            160 \\ 
\hline
   27 &       16 &       16 &       16 &       16 &       16 &              16 \\ 
\hline
   28 &      256 &      208 &      224 &      208 &      240 &            224 \\ 
\hline
   29 &        8 &        4 &        4 &        4 &        4 &                4 \\ 
\hline
   31 &     2048 &       69 &      225 &       73 &      242 &            229 \\ 
\hline
   32 &     1024 &      972 &     1008 &      972 &     1024 &          1008 \\ 
\hline
   33 &       64 &       22 &       24 &       22 &       24 &              24 \\ 
\hline
   34 &       16 &       16 &       16 &       16 &       16 &              16 \\ 
\hline
   36 &     4096 &     2308 &     2936 &     3084 &     3196 &          3280 \\ 
\hline
   37 &        4 &        4 &        4 &        4 &        4 &               4 \\ 
\hline
   38 &       64 &       24 &       24 &       24 &       24 &              24 \\ 
\hline
   39 &     2048 &      244 &      423 &      267 &      429 &            427 \\ 
\hline
   41 &        8 &        4 &        4 &        4 &        4 &               4 \\ 
\hline
   42 &     1024 &      504 &      702 &      530 &      706 &            702 \\ 
\hline
   43 &        4 &        4 &        4 &        4 &        4 &                4 \\ 
\hline
   44 &     4096 &     1484 &     1696 &     1692 &     1716 &          1840 \\ 
\hline
   46 &       16 &       16 &       16 &       16 &       16 &              16 \\ 
\hline
   47 &        4 &        4 &        4 &        4 &        4 &                4 \\ 
\hline
   48 & 1048576  &   400240 &   561252 &   445932 &   572536 &           579044 \\
\hline
   49 &        8 &        8 &        8 &        8 &        8 &                8 \\
\hline
   51 &       32 &       18 &       20 &       18 &       20 &              20 \\
\hline
   52 &    65536 &    13265 &    18552 &    15032 &    18676 &        19160 \\
\hline
   53 &        4 &        4 &        4 &        4 &        4 &                4 \\
\hline
   54 &      256 &      256 &      256 &      256 &      256 &            256 \\
\hline
    56 &    16384 &     6780 &     8396 &     7428 &     8824 &          8788 \\
\hline
    57 &       64 &       22 &       24 &       22 &       24 &              24 \\
\hline
    58 &       64 &       28 &       32 &       28 &       32 &              32 \\ 
\hline
 59 &        8 &        4 &        4 &        4 &        4 &         4 \\
 \hline
 61 &        8 &        4 &        4 &        4 &        4 &         4 \\
 \hline
\end{array}
$
\end{tiny}

\captionof{table}{\label{tab6}Tightness $\FF_5$, $8\leq n\leq 61$}
\end{minipage}
\end{figure}

\clearpage


\bibliography{RefsCGC}

\end{document}